\documentclass[aoas,nameyear,seceqn,dvips]{arximspdf}
\usepackage{graphicx}

\doi{10.1214/10-AOAS363}
\volume{4}
\issue{4}
\pubyear{2010}
\firstpage{1660}
\lastpage{1697}

\makeatletter

\newtheorem{theorem}{Theorem}[section]
\newtheorem{cor}[theorem]{Corollary}
\newproclaim{alg}[theorem]{Algorithm}

\def\eqref#1{(\ref{#1})}
\makeatother

\begin{document}
\begin{frontmatter}

\title{Subsampling Methods for genomic inference\thanksref{T1}}
\runtitle{Subsampling methods for genomic inference}

\thankstext{T1}{Supported by NIH U01 HG004695 (ENCODE DAC) and NIH
5R01GM075312.}
\begin{aug}
\author[A]{\fnms{Peter J.} \snm{Bickel}\corref{}\thanksref{t1}\ead[label=e1]{bickel@stat.berkeley.edu}},
\author[A]{\fnms{Nathan} \snm{Boley}\thanksref{t1}\ead[label=e2]{npboley@gmail.com}},
\author[A]{\fnms{James B.} \snm{Brown}\thanksref{t1}\ead[label=e3]{benbrownofberkeley@gmail.com}},
\author[A]{\fnms{Haiyan}~\snm{Huang}\thanksref{t1}\ead[label=e4]{hhuang@stat.berkeley.edu}}
\and
\author[B]{\fnms{Nancy R.} \snm{Zhang}\thanksref{t1}\ead[label=e5]{nzhang@stanford.edu}}

\thankstext[*]{t1}{The authors are ordered alphabetically.}
\runauthor{P. J. Bickel et al.}

\affiliation{University of California at Berkeley,
University of California at Berkeley,
University of California at Berkeley,
University of California at Berkeley, and~Stanford University}
\address[A]{P. J. Bickel\\
N. Boley\\
J. B. Brown\\
H. Huang\\
University of California at Berkeley\\
Berkeley, California\\
USA\\
\printead{e1}\\
\phantom{E-mail: }\printead*{e2}\\
\phantom{E-mail: }\printead*{e3}\\
\phantom{E-mail: }\printead*{e4}} %adresu isvedimo komanda gale!
\address[B]{N. R. Zhang\\
Stanford University\\
Stanford, California\\
USA\\
\printead{e5}}
\end{aug}

% HISTORY:
\received{\smonth{7} \syear{2009}}
\revised{\smonth{3} \syear{2010}}

% ABSTRACT
%
\begin{abstract}
Large-scale statistical analysis of data sets associated with genome
sequences plays an important role in modern biology. A key
component of such statistical analyses is the computation of
$p$-values and confidence bounds for statistics defined on the genome.
Currently such computation is commonly achieved through ad hoc
simulation measures. The method of randomization, which is at the
heart of these simulation procedures, can significantly affect the
resulting statistical conclusions. Most simulation schemes
introduce a variety of hidden assumptions regarding the nature of
the randomness in the data, resulting in a failure to capture
biologically meaningful relationships. To address the need for a
method of assessing the significance of observations within large
scale genomic studies, where there often exists a complex dependency
structure between observations, we propose a unified solution built
upon a data subsampling approach. We propose a piecewise
stationary model for genome sequences and show that the subsampling
approach gives correct answers under this model. We illustrate the
method on three simulation studies and two real data examples.
\end{abstract}

% KEYWORDS
%

\end{frontmatter}

%s1 ###
\section{Introduction}\label{sec1}
%s1.1 ###
\subsection{Background}\label{sec1.1}

This paper grew out of a number of examples arising in data coming from
the Encyclopedia of DNA Elements (ENCODE) Pilot Project [Birney et al.
(\citeyear{2007birney})], which is composed of multiple, diverse experiments performed
on a targeted 1\% of the human genome. Computational analyses of this
data are aimed at revealing new insights about how the information
coded in the DNA blueprint is turned into functioning systems in the
living cell. Variations of some of the methods described here have been
applied at various places in that paper, as well as in Margulies et al.
(\citeyear{2007margulies}), for assessing significance and computing confidence bounds for
statistics that operate along a genomic sequence. The background of
these methods is described in cookbook form in the supplements to those
papers, and it is the goal of this paper to present them rigorously and
to develop some
necessary refinements.

Essentially, we will argue that, in making inference about
statistics computed from ``large'' stretches of the genome, in the
absence of real knowledge about the evolutionary path which led to
the genome in question, the best we can do is to model the genome
by a piecewise stationary ergodic random process. The
variables of this process can be base pair composition or some other
local features, such as various annotated functional elements.
%
%This type of model has attracted increasing attention in recent
%years from the point of view of signal processiong and financial
%time series, e.g. Davis (2006) and forthcoming work of van Bellegen
%and Spokoiny and his students (2007).

In the purely stationary case some of the types of questions that we
will address, such as tests for independence of point processes,
confidence bounds for expectations of local functions and goodness of
fit of models, have been considered extensively. However,
inference for piecewise stationary models appears not to have been
investigated. With the advent of enormous amounts of genomic data,
all sorts of inferential questions have arisen. The proposed model
may be the only truly nonparametric approach to the genome, although,
just as in ordinary nonparametric statistics, there are many possible
ways of carrying out inference.

Our methods are based on a development of the resampling schemes of
Politis and Romano (\citeyear{1994politis}), Politis, Romano and Wolf (\citeyear{1999politis}) and the
block bootstrap methods of K\"{u}nsch (\citeyear{1989kunsch}). As we shall see, in
many situations, Gaussian approximations can replace these
schemes. But in these situations, as with the ordinary bootstrap,
we believe that a subsampling approach is valuable for the
following reasons:

\begin{itemize}
\item Letting the computer do the approximation is much easier.
\item Some statistics, such as tests of the Kolmogorov--Smirnov
type, are functions of stochastic processes to which a joint
Gaussian approximation applies. Then, limiting distributions can
only be computed by simulation.
\item The bootstrap distributions of our statistics show us
whether the approximate Gaussianity we have invoked for the ``true''
distribution of these statistics is in fact warranted. This visual
confirmation is invaluable.
\end{itemize}

This paper is organized as follows. We begin with some concrete
examples from the ENCODE data as well as other types of genomic data
in Section \ref{secexamples}, and proceed with a motivated
description of our model in Section \ref{secmodel}. Our methods are
discussed both qualitatively and mathematically in Sections
\ref{seclinearstat} and \ref{secsubsampling}. Section \ref{secsim}
contains results from simulation studies and real data analysis.
Proofs of theorems stated in Sections
\ref{seclinearstat} and \ref{secsubsampling} can be found in the
supplemental article [Bickel et al. (\citeyear{2010bickel-aoas})].

The statistics and methods discussed in this paper have been
implemented in several computing languages and are available for
download at \url{http://www.encodestatistics.org/}. Each of these implementations
runs in $n\log(n)$ time, where $n$ is the number of instances of
the more frequent feature. On a desktop PC (Intel Core Duo 3~GHz and
2~Gb RAM) the Python version takes over 1000 samples per second for
features on the order of $10^4$ instances.

%s1.2 ###
\subsection{Motivating examples} \label{secexamples}

We start with several fundamental questions that arise in genomic
studies.
\begin{itemize}
\item{\it Association of functional elements in
genomes.} In genomic analyses, a natural quantity of interest is the
association among different functional sites/features annotated
along the DNA sequence. Its biological motivation comes from the
common belief that significant physical overlapping or proximity of functional
sites in the genome suggests biological constraints or
relationships. In the ENCODE project, to understand the possible
functional roles of the evolutionarily constrained sequences that
are conserved across multiple species, overlap between the
constrained sequences and several experimental annotations, such as
5'UTR, RxFrags, pseudogenes, and coding sequences (CDSs), have
been evaluated using the method discussed in this paper. It was
found that the overlap of most experimental annotations with the
constrained sequences are significantly different from random
[Birney et al. (\citeyear{2007birney})]. An illustrative example from The ENCODE
Project [Birney et al. (\citeyear{2007birney})] is detailed in
Section~\ref{sec5.1}.
\item{\it Cooperativity between transcription factor binding sites.}
In some situations, there is interest to study the associations
between neighboring functional sites that do not necessarily
overlap. For instance, it is known that transcription factors often
work cooperatively and their binding sites (TFBS) tend
to occur in clusters [Zhang et al. (\citeyear{2006zhangM})].
Consequently, methods for identifying interacting
transcription factors usually involve evaluating the
significance of co-occurrences of their binding sites in a local
genomic region [Zhou and Wong (\citeyear{2004zhouPNAS}); Das, Banerjee and Zhang (\citeyear{2004das});
Yu, Yoo and Greenwald (\citeyear{2004yu}); Huang et al. (\citeyear{2004huang});
Kato et al. (\citeyear{2004kato}); Gupta and Liu (\citeyear{2005gupta})].
This problem has the same formulation as
the above ENCODE examples given a functional site defined as
follows: for a TFBS of length $l$ at position $i$, we define the
region $(i-m, i+l+m)$ as a functional site. Then two overlapping
functional sites are equivalent to two neighboring TFBSs with
interdistance less than $2m$, and the methods discussed in this
paper for evaluating the significance of overlapping functional
features can be applied. We leave this and related applications which
involve considering statistics of the K-S type to a later paper.
\item{\it Correlating DNA copy number with genomic content.} Recent
technology has made it possible to assay DNA copy number variation
at a very fine scale along the genome [for review, see Carter
(\citeyear{2007carter})]. Many studies, for example, Redon et al. (\citeyear{2006redon}), have shown that
such variation in DNA copy number is a common type of polymorphism
in the human genome. To what extent do these regions of copy number
changes overlap with known genomic features, such as coding
sequences? Redon et al. performed such an analysis and argued that
copy number variant regions have a significant paucity for coding
regions. The $p$-values supporting this claim were based on random
permutations of the start locations of the variant segments. This
assumes uniformity and stationarity of the copy number variants.
However, CNVs do not occur at random and are often clustered in regions
of the genome containing segmental duplications. The
methods discussed in this paper for evaluating the significance of
overlapping features, which assume neither uniformity nor
stationarity, can again be applied to this problem. Actually, the
results from our method suggest a different conclusion on this
problem (see Section \ref{real2}).

\end{itemize}

As we have seen in these examples, a common question asked in many
applications is the following: Given the position vectors of two
features in the genome, for example, ``conservation between species'' and
``transcription start sites,'' and a measure of relatedness between
features, for example, base or region percentage overlap, how
significant is
the observed value of the measure? How does it compare with that
which might be observed ``at random?''

The essential challenge in the statistical formulation of this
problem is the appropriate modeling of randomness of the genome,
since we observe only one of the multitudes of possible genomes that
evolution might have produced for our and other species.

How have such questions been answered previously? Existing methods
employ varied ways to simulate the locations of features within
genomes, but all center around the uniformity assumption of the
features' start positions: The features must occur homogeneously in
the studied genome region, for example, Blakesley et al. (\citeyear{2004blakesley}) and
Redon et
al. (\citeyear{2006redon}). This assumption ignores the natural clumping of features
as well as the nonstationarity of genome sequences. Clumping of
features is quite common along the genome due to either the
feature's own characteristic, for example, transcription factor binding
sites (TFBSs) tend to occur in clusters, or the genome's
evolutionary constraints, for example, conserved elements are often
found in
dense conservation neighborhoods. Ignoring these natural properties
could result in misleading conclusions.

%On the basis of a more biologically meaningful elaboration of the
%randomness of the features' start positions, this paper introduces a
%reliable method for evaluating the feature relationships that are
%defined through linear statistics.
%We propose a model of the features along the genome which we view as
%``nonparametric" as possible. Its biological motivation is
%plausible, as we partially demonstrate in several real examples.

In this paper we suggest a piecewise stationary model for the
genome (see Section \ref{secmodel}) and, based on it, propose a
method to infer the relationships between features which we view as
``nonparametric'' as possible (see Sections \ref{SegBlock} and
\ref{hypotesting}). The model is based on assumptions which we
demonstrate in real data examples to be
plausible.

%s2 ###
\section{The piecewise stationary model} \label{secmodel}
%s2.1 ###
\subsection{Genomic motivation}\label{sec2.1}
We postulate the following for the observed ge\-nomes or genomic
regions:
\begin{itemize}
\item They can be thought of as a
concatenation of a number of regions, each of which is homogenous in
a way we describe below.
\item Features that are located very far
from each other on the average have little to do with each other.
\item The number of such homogeneous regions is small compared to
the total length of the observed genome.
\end{itemize}
These assumptions, which form the underpinning of
our \emph{block stationary model} for genomic features, are
motivated by earlier studies of DNA sequences, which show that there
are global shifts in base composition, but that certain sequence
characteristics are locally unchanging. One such characteristic is
the GC content. Bernardi et al. (\citeyear{1985bernardi}) coined the term
``isochore''
to denote large segments (of length greater than 300 Kb) that have
fairly homogeneous base composition and, especially, constant GC
composition. Even earlier, evidence of segmental DNA structure can
be found in chromosomal banding in polytene chromosomes in
drosophila, visible through the microscope, that result from
underlying physical and chemical structure. These banding patterns
are stable enough to be used for the identification of chromosomes
and for genetic mapping, and are physical evidence for a block
stationarity model for the GC content of the genome.

The experimental evidence for segmental genome structure and the
increasing availability of DNA sequence data have inspired attempts
to computationally segment DNA into statistically homogeneous
regions. The paper by Braun and M\"{u}ller (\citeyear{1998braun}) offers a
review of statistical methods developed for detecting and modeling
the inhomogeneity in DNA sequences. There have been many attempts to
segment DNA sequences by both base composition [Fu and
Curnow (\citeyear{1990fu}); Churchill (\citeyear{1989churchill}, \citeyear{1992churchill}); Li et al. (\citeyear{2002li})] and chemical
characteristics [Li et al. (\citeyear{1998li})]. Most of these computational
studies concluded that a model that assumes block-wise stationarity
gives a significantly better fit to the data than stationary models
[see, for example, the conclusions of two very different studies by
Fickett, Torney and Wolf (\citeyear{1992fickett}) and Li et al. (\citeyear{1998li})].

A subtle issue in the definition of ``homogeneity'' is the scale at
which the genome is being analyzed. Inhomogeneity at the kilobase
resolution, for example, might be ``smoothed out'' in an analysis at the
megabase level. The level of resolution is a modeling issue that
must be considered carefully with the goal of the analysis in mind.

Implicit in our formulation is an ``ergodic'' hypothesis. We want
probabilities to refer to the population of potential genomes. We
assume that the statistics of the genome we have mimic those of the
population of genomes. This is entirely analogous to the ergodic
hypothesis that long term time averages agree with space averages
for trajectories of dynamic systems.

%s2.2 ###
\subsection{Mathematical formulation}\label{sec2.2}
In mathematical terms, the block stationarity model assumes that we
observe a sequence of random variables $\{X_1,\dots, X_n\}$
positioned linearly along the genomic region of interest. $X_k,
k=1,\ldots,n,$ may be base composition, or some other measurable
feature. We assume that there exist integers $\bolds{\tau} =
\bolds{\tau}^{(n)} =(\tau_0,\dots,\tau_U)$, where
$0=\tau_0<\tau_1<\cdots<\tau_U=n$, such that the collections of
variables, $\{X_{\tau_i},\ldots, X_{\tau_{i+1}}\}$, are separately
stationary for each $i=0,\dots,U-1$. We let $n_i =
\tau_i-\tau_{i-1}$ be the length of the $i$th region, and let there
be $U$ such regions in total. For convenience, we introduce the
mapping
\[
\pi\dvtx \{1,\dots, n\} \rightarrow\{(i,j)\dvtx 1\leq i \leq U, 1\leq j
\leq n_i\}
\]
which relates the relabeled sequence, $\{X_{ij}\dvtx 1\leq
i \leq U, 1\leq j \leq n_i\}$, to the original sequence
$\{X_1,\ldots,X_n\}$. We write $\pi=(\pi_1,\pi_2)$ with $\pi(k) =
(i,j)$ if and only if $k = \tau_i+j$.
We will use the notation $X_{ij}$ and $X_k$ interchangeably throughout
this paper.

For any $k_1$, $k_2$, let $\mathcal{F}_{k_1}^{k_2}$ be the
$\sigma$-field generated by $X_{k_1},\dots, X_{k_2}$. Define $m(k)$
to be the standard Rosenblatt mixing number [Dedecker et al. (\citeyear{2007dedecker})],
\[
m(k) = \sup\{|\mathbb{P}(AB)-\mathbb{P}(A) \mathbb{P}(B)|\dvtx  A \in
\mathcal
{F}_1^l, B \in\mathcal{F}_{l+k}^n,  1\leq l \leq n-k\}.
\]
Then, assumptions 1--3 stated in Section~\ref{sec2.1}
translate to the following:

\begin{enumerate}[A1.]
\item[A1.] The sequence $\{X_1,\ldots,X_n\}$ is piecewise stationary.
That is, $\{X_{ij}\dvtx 1\leq j\leq n_i\}$ is a stationary sequence for
$i=1,\dots,U$.
\item[A2.] There exists constants $c$ and $\beta>0$ such that
$m(k)\leq ck^{-\beta}$ for all $k$.
\item[A3.] $U/n \rightarrow0$.
\end{enumerate}

An immediate and important consequence of A1--A3 is that, for any
fixed small $k$, if we define $W_1=(X_1,\ldots,X_k),
W_2=(X_{k+1},\ldots, X_{2k}),\ldots, W_m=(X_{n-k+1},\ldots, X_n)$,
where $m=n/k$, then $\{W_1,\ldots, W_m\}$ also obey A1--A3. This is
useful, for example, in the region overlap example considered in the
next section.

The remarkable feature of these assumptions, which are more general
to our knowledge than any made heretofore in this context, is that
they still allow us to conduct most of the statistical inference of
interest. Not surprisingly, these assumptions lead to more
conservative estimates of significance than any of the previous
methods.

%s3 ###
\section{Vector linear statistics and Gaussian approximation} \label{seclinearstat}

We study the distribution of a class of vector linear statistics of
interest under the above piecewise stationary model. As an
illustration, we consider the ENCODE data examples, and suppose that
we are interested in base pair overlap between features $A$ and
$B$. We can represent base pair overlap by defining
\begin{eqnarray*}
&&I_k=1,\qquad \mbox{if position $k$ belongs to feature $A$ and $0$ otherwise},
\\[-2pt]
&&J_k=1,\quad\hspace*{10pt} \mbox{if position $k$ belongs to feature $B$ and $0$ otherwise}.
\end{eqnarray*}
We can then define $X_k = I_kJ_k$ to be the indicator
that position $k$ belongs to both features $A$ and  $B$. Then,
for the $n=30$ megabases of the ENCODE regions, the mean base pair
overlap is equal to
\[
\bar{X}=\sum_{k=1}^n X_k/n.
\]

Another biologically interesting statistic is the (asymmetric)
region overlap, defined as follows: suppose the consecutive feature
stretches are $T_1,\ldots,T_{\alpha}$ with lengths
$\tau_1,\ldots,\tau_{\alpha}$, and the corresponding nonfeature
stretches $S_1,\ldots,S_{\beta}$ with lengths
$\rho_1,\ldots,\rho_{\beta}$. We assume here that the initial and final
stretches consist of one feature and one nonfeature stretch. The
complementary situation, when both initial and final stretches are
of the same type, is dealt with similarly. Without loss of
generality, suppose the initial stretch is nonfeature. Then, $S_1 =
\{1,\ldots,\rho_1\}$, $T_1 = \{\rho_1 + 1,\ldots,\rho_1+\tau_1\}$, $S_2 =
\{\rho_1+\tau_1+1,\ldots, \rho_1+\tau_1+\rho_2\}$, etc. Using $I_k$,
$J_k$ as indicators of feature identity, we define the
(unnormalized) region overlap of feature~$A$ stretches with feature $B$
stretches as $\frac{1}{n} \sum_{t=1}^{\alpha} V_t $ where $V_t = 1 -
\prod_{k \in T_{A,t} } (1 - J_k)$, where $T_{A,1},\ldots,T_{A,\alpha}$
denote the feature~$A$ stretches. This statistic is not linear in
terms of functions of single basepairs, but is linear in functions
of blocks of feature $B$. These blocks are of random sizes, but
are consistent with our hypothesis of piecewise stationarity that,
except for end effects due to feature instances crossing segment
boundaries, the ${V_t}$ are also stationary. If the lengths
$\tau_1,\ldots,\tau_{\alpha}$ are negligible compared to $n$ and
$\alpha$ is of the order of $n$, we can expect the mixing
hypothesis to remain valid.

In general, we focus our attention
on statistics that can be expressed as a function of the mean of
$\mathbf{g}(X_i)$, where $\mathbf{g}$ is some well behaved
$d$-dimensional vector
function to be
characterized in later sections. By the flexible definition of
$\mathbf{g}$, this encompasses a wide class of situations.

First, we consider vector linear statistics of the form
\[
\mathbf{T}_n(\mathbf{X})=n^{-1}\sum_{k=1}^n\mathbf{g}(X_k).
\]
We introduce the following
notation:
\begin{eqnarray*}
E[\mathbf{T}_n] \equiv \bolds{\mu}\equiv\sum_{i=1}^U f_i\bolds
{\mu}_i,
\end{eqnarray*}
where
\begin{eqnarray*}
\bolds{\mu}_i &\equiv& E[\mathbf{g}(X_{i1})] ,
\\
f_i &\equiv& n_i/n
\end{eqnarray*}
and
%
%e3.1 ###
\begin{equation}\label{eq:Sigma}
\Sigma_n \equiv\operatorname
{\mathbb{V}ar}(
n^{{1/2}} \mathbf{T}_n) = \sum^U_{i=1} f_i
C_i(nf_i),
\end{equation}
where
%
%e3.2 ###
\[
C_i(m) =
C_{i0} + 2 \sum^m_{\ell=1} C_{i\ell} \biggl( 1 - \frac{(\ell-1)}{m}
\biggr)
\]
and
%e3.3 ###
\begin{equation}\label{eq:C}
C_{i0} \equiv\operatorname{\mathbb{V}ar}\mathbf{g}(X_1), \qquad
C_{i\ell} \equiv
\operatorname{\mathbb{C}ov}\bigl( \mathbf{g}(X_{i1}), \mathbf
{g}\bigl(X_{i(l+1)}\bigr) \bigr).
\end{equation}
In
Theorem \ref{theorem:linear} below, we show asymptotic Gaussianity of
$\mathbf{T}_n$ given a few more technical assumptions:
\begin{enumerate}[A4.]
\item[A4.] $\frac{1}{n} \sum_{i: n_i \leq l} n_i
\rightarrow0$ for all $l< \infty$.
\item[A5.] $\forall i$, $|\mathbf{g}|_{\infty} \leq C < \infty$.
%%TODO:
%define g
%
\item[A6.] $ 0<\varepsilon_0 \leq\|\Sigma_n \| \leq\varepsilon
_0^{-1} $, for all
$n$, where $\| \cdot\|$ is a matrix norm.
\end{enumerate}
In
particular, A4 implies that the contribution of ``small regions'' to
the overall statistic must not be too large.

\begin{theorem}
\label{theorem:linear} Under conditions \textup{A1--A6},
%
%e3.4 ###
\begin{equation} \label{eqL1} n^{{1/2}}
\Sigma_n^{-{1/2}} (\mathbf{T}_n -\bolds{\mu}) \Rightarrow
\mathcal N(\mathbf{0},\mathbf{I}),
\end{equation}
where $\mathbf{I}$ is the $d \times d$ identity.
\end{theorem}

The proof of the theorem is in the supplemental article [Bickel et al. (\citeyear{2010bickel-aoas})].
If we have estimates $\hat{\bolds{\tau}}$ of $\bolds{\tau}$ which are
consistent in a suitably uniform sense, then estimates of
$C_{i\ell}$, $C_{i}(m)$ using $\hat{\bolds{\tau}}$ in place of
$\bolds{\tau}$ are also consistent. However, simply plugging these
estimates into (\ref{eq:Sigma}) does not yield consistent estimates
of $\sigma^2$ if our approach were to compute confidence intervals
by Gaussian approximation. This is well known for the stationary
case. Some regularization is necessary. We do not pursue this
direction but prefer to approach the inference problem from a
resampling point of view---see next section.

In many cases, the statistics of interest are not linear. For
example, in the analysis of the ENCODE data a more informative
statistic is the \%bp overlap defined as
%
%e3.5 ###
\begin{equation}\label{eq:6}
B \equiv\frac{\bar{X}}{D},
\end{equation}
where
\begin{eqnarray*} D=\sum_{k=1}^nI_k
\end{eqnarray*}
is the total
base count of feature $A$.

More conceptually, the region overlap is
%
%e3.6 ###
\begin{equation}\label{eq:7}
 R \equiv
\frac{1}{W_I} \sum_{k=1}^{K}V_k,
\end{equation}
where
$W_I = \sum_{k=1}^{n}I_{k-1}(1-I_k)$, the number of feature~$A$
instances.

A standard delta method computation shows that the standard error of
$B$ can be approximated as follows: Let $\mu(D)$ and $\mu(\bar{X})$
be respectively the expectation of $D$ and $\bar{X}$. Then,
\begin{eqnarray*} \frac{\bar{X}}{D}-\frac{\mu(\bar{X})}{\mu
(D)}\approx
\frac{\bar{X}-\mu(\bar{X})}{\mu(D)}-\mu(\bar{X})\frac{(D-\mu
(D))}{\mu^2(D)},
\end{eqnarray*}
and, hence, we can approximate $\frac{\bar{X}}{D}$ by a
Gaussian variable with mean $\frac{\mu(\bar{X})}{\mu(D)}$ and variance
%
%e3.7 ###
\begin{equation}%\label{eq:7}
 \sigma^2(B)\approx
\frac{\sigma^2(\bar{X})}{\mu^2(D)}+\frac{\mu^2(\bar{X})}{\mu
^4(D)}\sigma^2(D)
- 2\frac{\mu(\bar{X})}{\mu^3(D)}\operatorname{Cov}(\bar{X},D),
\end{equation}
where $\sigma^2(B)$, $\sigma^2(\bar{X})$, $\sigma^2(D)$
are the corresponding variances and $\operatorname{Cov}(\bar{X},D)$ denotes the
covariance. In doing inference, we can use the approximate
Gaussianity of $B$ with $\sigma^2(B)$ estimated using the above
formula with regularized sample moments replacing the true moments.

We also note that goodness of fit or equality of population test
statistics, such as Kolmogorov--Smirnov and many others, can be
viewed as functions of empirical distributions, which themselves are
infinite-dimensional linear statistics, and we expect, but have not
proved, that the
methods discussed in this paper and the underlying theories apply to
those cases as well, under suitable assumptions.

%s4 ###
\section{Subsampling based methods} \label{secsubsampling}
Here we propose a subsampling based approach, in particular, a
combined segmentation-block subsampling method to conduct statistical
inference under the piecewise stationary model, which we call
``segmented block subsampling.'' In our method, the segmentation
parameters governing scale are chosen first and then the size of the
subsample is chosen based on stability criteria. The segmentation
procedure, as we discussed, is motivated by the heterogeneity of
large-scale genomic sequences. The block subsampling
approach takes into account the local genomic structure, such
as natural clumping of features, when conducting statistical
inference. We explicitly demonstrate the advantages of using
segmentation and block subsampling by simulation studies in Section
\ref{secsim}.

%s4.1 ###
\subsection{Stationary block subsampling}\label{sec4.1}

Below we first review the results related to the stationary
block bootstrap method in a homogeneous region ($U=1$), and then show
how the method breaks down when it is applied to a piecewise stationary
sequence ($U>1$).

%s4.1.1 ###
\subsubsection{Review of results for the case of $U=1$} \label{sec:stationary}

For completeness, we recall the following basic algorithm of Politis
and Romano (\citeyear{1994politis}) to obtain an estimate of the distribution of the
statistic $\mathbf{T}_n(X_1,\ldots,X_n)$ under the assumption that the
sequence $X_1,\ldots,X_n$ is stationary (i.e., $U=1$).

\begin{alg}
\label{alg:4:0}
\textup{(a)} Given $L \ll n$ choose a number $N$ uniformly at
random from $\{1,\ldots, n-L\}$.\vspace*{-6pt}
\begin{longlist}
\item[(b)] Given the statistic $\mathbf{T}$,
as above, compute
\[
\mathbf{T}_L (X_{N+1},\ldots,X_{N+L}) \equiv\mathbf{T}_{L1}^*.
\]
\item[(c)] Repeat $B$ times with replacement to obtain
$\mathbf{T}_{L1}^*,\ldots, \mathbf{T}_{LB}^*$.
\item[(d)] Estimate the distribution
of $\sqrt{n}(\mathbf{T}_n- \mu)$ by the empirical distribution
$\mathcal L_B^*$
of
\[
\Biggl\{ \sqrt{\frac{n}{L} } [ \mathbf{T}_{Lj}^* - \mathbf
{T}_n(X_1,\ldots,X_n)
],  1 \leq b \leq B \Biggr\}.
\]
\end{longlist}
\end{alg}

By Theorem 4.2.1 of Politis, Romano and Wolf (\citeyear{1999politis}),
%
%e4.1 ###
\begin{equation}\label{eqT2}
\mathcal L_B^* \Longrightarrow\mathcal N_d(\mathbf{0} ,\Sigma)
\end{equation}
in probability
for some constant $\Sigma$ if \eqref{eqL1} holds and if $L\rightarrow
\infty$, $L/n \to0$. As usual,
convergence of $\mathcal L_B^*$ in law in probability simply means that
if $\rho$ is any metric for weak convergence on $R^d$, then $\rho(
\mathcal L_B^*, \mathcal L) \mathop{\rightarrow}\limits^P 0$.

Since all variables we deal with are in $L_2$, we take $\rho$ to be
the Mallows metric,
\[
\rho_M^2 (F,G) = \min\{ E_P (W-V)^2\dvtx  P
\mbox{ such that } W \sim F,  V \sim G \}  .
\]
Useful
properties of $\rho_M$ are as follows:
\begin{longlist}
\item[(a)] $\rho_M^2
( \Sigma\pi_i F_i, \Sigma\pi_i G_i ) \leq\Sigma\pi_i
\rho_M^2 (F_i,G_i) $ for all $ \pi_i \geq0,  \Sigma\pi_i =1 $ and
\item[(b)] If $F=F_1 * \cdots* F_m$, $G=G_1 * \cdots* G_m$, that is,
$F$ and $G$ are distributions of sums of $m$ independent variables,
then $\rho_M^2 (F,G) \leq\sum^m_{i=1} \rho_M^2
(F_i,G_i)$.
\end{longlist}
For convenience, when no confusion is possible, we will
write $\rho_M(W,V)$ for $\rho_M(F,G)$ for random variables $W\sim
F$, $V\sim G$.

%We cite the specialization of a theorem (essentially 4.4.1) of
%Politis, Romano and Wolf (1999) justifying \eqref{eqT1} for linear
%statistics. \begin{theorem} \label{theorem:4:1}Suppose (A1)-(A3) hold and
%$\bmg: R^d \to R$ with $E\bmg(X_1) = \0$ and $|\bmg|_{\infty} \leq c
%< \infty$. Then \be\label{eqT3} n^{- \frac{1}{2} } \sum^n_{i=1}
%(X_i), \bmg(X_{1+i}) ) \ ,$$ where, as usual, if $E\mathbf{U} = \0$
%$$\Var(\mathbf{U})=E\mathbf{U}\mathbf{U}^T,\ \
%Politis, Romano
%and Wolf require weaker mixing conditions than (A2) and only $2+
%conditions apply to the results for the segmented case we present
%but we do not pursue this.

%s4.1.2 ###
\subsubsection{Performance of the block subsampling method in the
piecewise stationary model when $U>1$} \label{secinconsist}

We turn to the analogue of Theorem 4.2.1 in Politis, Romano and Wolf
(\citeyear{1999politis}) for $U>1$. We consider a vector linear statistic, for which
the true distribution was described in Section \ref{seclinearstat}.
Here, we ask how Algorithm \ref{alg:4:0}, which assumes
stationarity, performs in this nonstationary context. We show that,
in general, it does not give correct confidence bounds but is
conservative, sometimes exceedingly so. The results depend on $L$,
the subsample size, which is a crucial parameter in Algorithm \ref{alg:4:0}.
We sketch these issues in Theorem \ref{theorem:4:3} below, for
simplicity, letting $g$ be the one-dimensional identity function
$g(x)=x$. Let
\begin{eqnarray*}
\tau^2 &=& U^{-1} \sum^U_{i=1} (\mu_i - \bar{\mu})^2,
\\
\bar{X}_i &\equiv& n_i^{-1} \sum^{n_i}_{j=1} X_{ij},\qquad
\bar{X}\equiv n^{-1} \sum^n_{k=1} X_k = \sum^U_{i=1} f_i \bar{X}_i.
\end{eqnarray*}
Also let
\[
n_i^* \equiv\mbox{ Cardinality of }
S_i\equiv\{ k \dvtx  k \in[N,N+L],  \pi_1(k)=i \}
\]
and
\begin{eqnarray*}
\bar{X}_i^* &=& 1(n_i^* \neq0) \sum_j \{ X_{ij}\dvtx  j \in S_i \}
/ n_i^*  ,
\\
\bar{X}_L^* &=& \sum^U_{i=1} f_i^* \bar{X}_i^*
\qquad\mbox{where } f_i^* \equiv\frac{ n_i^* }{L}.
\end{eqnarray*}

We introduce one assumption that is obviously needed for any
analysis of the block or segmented resampling bootstraps:
\begin{enumerate}[A7.]
\item[A7.] $L\rightarrow\infty$,
\end{enumerate}
%
% and two more assumptions below, which are stronger than A4.
and two other assumptions which are used in different parts of
Theorem \ref{theorem:4:3} but not in the rest of the paper, and are thus
given a different numbering:
\begin{enumerate}[B1.]
\item[B1.] $L/n \rightarrow0 $.
\item[B2.] $(LU)/n \rightarrow0 $.
\end{enumerate}

\begin{theorem} \label{theorem:4:3} Let $\mathcal L_n$ be the distribution which
assigns mass $f_i$ to
$(\mu_i -\mu)$, $1 \leq i \leq U$, and write $C_i$ for $C_i(nf_i)$.
Suppose assumptions \textup{A1--A5} and \textup{A7} hold:
\begin{enumerate}[(iii)]
\item[(i)] If \textup{B2} holds, $\rho_M (\bar{X}_{L }^*- \bar{X}, \mathcal
L_n) \mathop{\longrightarrow}\limits^P 0 $.
\item[(ii)] If
%
%e4.2 ###
\begin{equation}
\label{eqT4.3.0} \sum^U_{i=1} f_i (\mu_i - \mu)^2 = o(L^{-1})
\end{equation}
and \textup{B1} holds, then
\[
\rho_M \Biggl[ \sqrt{L} ( \bar{X}_{L }^*- \bar{X}), \sum^U_{i=1} f_i
\mathcal N(0,C_i) \Biggr] \stackrel{P}{\longrightarrow} 0.
\]
\item[(iii)] If \eqref{eqT4.3.0} and \textup{B1} hold and
%
%e4.3 ###
\begin{equation}\label{condition2} \sum^U_{i=1} f_i 1 \bigl( |
\Sigma_n -C_i | \geq\varepsilon\bigr) \to0
\end{equation}
for all $\varepsilon>0$, then
\[
\rho_M \bigl( \sqrt{L} (
\bar{X}_{L}^*- \bar{X}), \mathcal N(0, \Sigma_n) \bigr)
\mathop{\longrightarrow}\limits^P 0.
\]
\end{enumerate}
\end{theorem}

The implications of Theorem \ref{theorem:4:3} are as follows. If equation
(\ref{eqT4.3.0}) does not hold, then $\bar{X}^*_L -\bar{X}$ does not
converge in law at scale $L^{-{1/2} }$ so that it does not
reflect the behavior of $L^{{1/2} }(\bar{X}_L -\mu)$ at all.
This is a consequence of inhomogeneity of the segment means.
Evidently in this case, confidence intervals of the percentile type
for $\mu$, $[ \bar{X} +c_n(\alpha), \bar{X} +c_n(1-\alpha
)],
$ where $c_n(\alpha)$ is the $\alpha$ quantile of the distribution
of $\bar{X}^*_L -\bar{X}$, will have coverage probability tending to
1, since $c_n(\alpha)$ and $c_n(1-\alpha)$ do not converge to 0 at
rate $L^{-{1/2}}$ as they should. If B2 does
not hold, we have to consider the possibility that [$N, N+L$] covers
$K_N$ consecutive segments, whose total length is $o(n)$, such that
the average over all such blocks is close to $\mu$. However, in the
absence of a condition such as (\ref{eqT4.3.0}) or mutual cancellation of
$\mu_i^*$, the scale of $\bar{X}_L^*$ will be larger than $L^{-1/2}$.
These issues will be clarified by the proof of Theorem~\ref{theorem:4:3} in the
supplemental article [Bickel et al. (\citeyear{2010bickel-aoas})]. We note also that (\ref{eqT4.3.0})
can be weakened to requiring that
the mean of blocks of consecutive segments whose total length is
small compared to $n$ be close to $\mu$ to order $o(L^{-1/2})$. But
our statement makes the issues clear. Finally, note that B2 holds
automatically if the number of segments $U$ is bounded and if B1
holds.

If (\ref{eqT4.3.0}) does hold but (\ref{condition2}) does not, then
$\sqrt{L}(\bar{X}^*_L -\bar{X})$ converges in law to the Gaussian
mixture $\sum^U_{i=1} f_i \mathcal N(0,C_i)$. The mixture of Gaussians is
more dispersed in a rough sense than a Gaussian with the same
variance, which is
\[
\sigma^2_n = \sum^U_{i=1} f_i C_i  ;
\]
see
Andrews and Mallows (\citeyear{1974andrews}). Especially note that, if $W$ has the
mixture distribution and $V$ is the Gaussian variable with the same
variance, then
\[
Ee^{tW} = \sum f_i e^{- {(t^2/2)} c_i} \geq
e^{- {t^2/2} \sum f_i C_i} = Ee^{tV}
\]
by Jensen's
inequality. This suggests that the tail probabilities will also be
overestimated. The overdispersion here, which leads to
conservativeness that is not as extreme as in case (i), is due to
inequality of the variances from segment to segment. Finally, if
(\ref{condition2}) holds, then the segments have essentially the same
mean and variance and stationary block subsampling works.

A mark of either (\ref{eqT4.3.0}) or (\ref{condition2}) failing is a
lack of Gaussianity in the distribution of $\bar{X}^*_L -\bar{X}$.
This was in fact observed at some scales in the ENCODE project,
which led us to crudely segment on biological grounds with reasonable
success. However, the correct solution, which we now present in this
paper, is to
estimate the segmentation and appropriately adjust the subsampling procedure.

%s4.2 ###
\subsection{A segmentation based block subsampling method}
\label{SegBlock}

We saw in the previous section that the na\"{i}ve block subsampling
method that was designed for the stationary case breaks down when
the sequence follows a piecewise stationary model. We propose a
stratified block subsampling strategy, which stratifies the subsample
based on a ``good'' segmentation of the sequence which is estimated
from the data. We first state the block subsampling method, and then
in Section \ref{secmainresult} give minimal conditions on the
estimated segmentation for its consistency. In Section \ref{secseg}
we discuss possible segmentation methods.

%s4.2.1 ###
\subsubsection{Description of algorithm}\label{secalg}

Assume that we are given a segmentation $\mathbf{t}=(0=t_0, t_1, \dots,
t_{m+1}=n)$, where $m$ is the number of regions in $\mathbf{t}$. Assume
that the total size $L$ of the subsample is pre-chosen. We define a
stratified block subsampling scheme as follows.

\begin{alg} \label{algsubsample}
For $i=1,\dots,m$, let $\lambda_i=\lambda_i(t) = \lceil
(t_i-t_{i-1})L/n \rceil$. We use the
notation $X_{i;l}$ to denote the block of length $l$ starting at $i$:
\[
X_{i;l} = (X_i,\dots,X_{i+l-1}).
\]
Then, for each subsample,

Draw integers $\mathbf{N}=\{N_1,\dots,N_m\}$, with $N_i$
chosen uniformly from $\{t_{i-1}+1,\dots,t_i-\lambda_i(t)+1\}$, and let
\begin{eqnarray*}
X^* = (X_1^*,\dots, X_m^*)= \bigl(X_{N_1;\lambda
_1(t)},\dots,X_{N_m;\lambda_{m}(t)}\bigr).
\end{eqnarray*}
 Repeat the above $B$ times to obtain $B$ subsamples:
$X^{*,1}, \dots, X^{*,B}$.
\end{alg}

To obtain a confidence interval for
$\bolds{\mu}$, we assume that the
statistic $\mathbf{T}_n$ has approximately a $N(\bolds{\mu},
\Sigma_n/n)$
distribution as in the previous section. For each subsample drawn as
described in Algorithm \ref{algsubsample}, compute the statistic
$\mathbf{T}_L^{*,b} = \mathbf{T}_L^{*,b}(\mathbf{t}) =\mathbf
{T}_L(X^{*,b})$. Form the sampling
estimate of variance,
%
%e4.4 ###
\begin{equation}\label{subsampvar}
\widehat{\Sigma}_n\equiv
\frac{L}{B}\sum_{b=1}^B(\mathbf{T}^{*,b}_L-\bar{\mathbf
{T}}^{*}_L)'(\mathbf{T}
^{*,b}_L-\bar{\mathbf{T}}^{*}_L),
\end{equation}
where $\bar{\mathbf{T}}^*_L \equiv\sum_{b=1}^B \mathbf
{T}_L^{*,b}/B$. We can now
proceed to estimate the confidence interval for $\mathbf{T}_n$
in standard ways. For example, in the univariate case where
$\sigma^2_n \equiv\Sigma_n$ is a scalar:
\begin{longlist}[(a)]
\item[(a)] Use $\bar{X}\pm
z_{1-\alpha/2}\frac{\widehat{\sigma}_n}{\sqrt{n}}$, where
$z_{1-\alpha/2}$
is the $1-\alpha/2$th quantile of $N(0,1)$, for a $1-\alpha$ confidence
interval.
\item[(b)] Efron's percentile method: Let
$\bar{X}_{(1)}^{*}<\cdots<\bar{X}_{(B)}^{*}$ be the ordered
$\bar{X}^{*,b}$, then use $[\bar{X}^{*}_{([B\alpha/2])},
\bar{X}^{*}_{([B(1-\alpha/2)])}]$ as a $1-\alpha$ confidence
interval.\vspace*{2pt}

\item[(c)] Use a Studentized interval [Efron (\citeyear{1981efron})] or
Efron's (\citeyear{1987thisted}) $\mathit{BCA}$ method; see Hall (\citeyear{1992hall}) for an extensive
discussion.
\end{longlist}

Although the theory for (c) giving the best coverage approximation
has not been written down, as it has been for the ordinary bootstrap, we
expect it to continue to hold. Evidently, these approaches can be
applied not only to vector linear statistics like $\mathbf{T}_n$ but
also to
smooth functions of vector linear statistics.

This algorithm assumes a given segmentation $\mathbf{t}$, which
should be set to some good estimate
$\hat{\bolds{\tau}}^{(n)}=\{0=\hat{\tau}_0,\hat{\tau
}_1,\ldots,\hat{\tau
}_m=n\}$ of
the true change points $\bolds{\tau}^{(n)}$. In order for the
algorithm to
perform well, a good segmentation is critical unless the sequence is
already reasonably homogeneous. In Section~\ref{secconsistrue} below we state the
result that the algorithm is consistent if the given segmentation
equals the true changepoints. Then, in Section~\ref{secmainresult}, we state a few
assumptions on the data determined segmentation $\hat{\bolds{\tau
}}^{(n)}$ which
would enable us to act as if the segmentation were known and state
Theorem~\ref{couplethm} to that effect.

%s4.2.2 ###
\subsubsection{Consistency with true segmentation}
\label{secconsistrue}

Under the hypothetical situation where the segmentation $\mathbf{t}$
assumed in Algorithm \ref{algsubsample} is equal to the set of true
changepoints,
then the algorithm can be easily shown to be consistent. Here we
state the result, which will be proved in the supplemental article
[Bickel et al. (\citeyear{2010bickel-aoas})].

First, we state a stronger version of the assumption B1, which
requires that the square of the subsample size $L=L_n$ be $o(n)$:
\begin{enumerate}[A8.]
\item[A8.] $ L_n^2/n \rightarrow0$.
\end{enumerate}
Then, the consistency of Algorithm
\ref{algsubsample} given the true segmentation follows from the
following theorem.

\begin{theorem} \label{theorem:trueconsistent} If
assumptions \textup{A1--A8} hold, then
%
%e4.5 ###
\begin{equation}
L_n^{1/2}{\Sigma_n}^{-1/2}[T^*_{L_n}(\tau_n) - T_n] \Rightarrow
N(0,I)
\end{equation}
in probability, where $I$ is the $d\times d$ identity.
\end{theorem}

%s4.2.3 ###
\subsubsection{Consistency with estimated segmentation}
\label{secmainresult}

Let
\[
\hat{\bolds{\tau}}=\hat{\bolds{\tau}}^{(n)}= \bigl(\hat{\tau
}^{(n)}_1,\ldots, \hat{\tau}^{(n)}_{\hat{U}_n}\bigr)
\]
be a
segmentation of the sequence $X_1,\dots, X_n$, which is determined
from the data, and which is intended to estimate the true
changepoints $\bolds{\tau}= \bolds{\tau}^{(n)}$. We will state
conditions on $\hat{\bolds{\tau}}$
such that the statistic obtained from Algorithm \ref{algsubsample}
based on $\hat{\bolds{\tau}}$ is close to the statistic obtained
from the same
algorithm based on the true segmentation $\bolds{\tau}$. This can be stated
formally as follows. For any segmentation $\mathbf{t}$, let
$\mathbf{X}^*(\mathbf{t})$ be a subsample drawn according to Algorithm
\ref{algsubsample} based on $\mathbf{t}$. Let $F^*_{n,\mathbf
{t}}(\cdot)$ be
the distribution of $\sqrt{L}\{T[\mathbf{X}^*(\mathbf{t})] -
E^*T[\mathbf{X}^*(\mathbf{t})]\}$ conditioned on $X_1,\dots,X_n$ and
$\mathbf{t}$. Then, the desired property on the estimated segmentation
$\hat{\bolds{\tau}}$ is that
%
%e4.6 ###
\begin{equation}\label{conv1}
\rho_M^2\bigl[F^*_{n,\hat{\bolds{\tau}}^{(n)}},
F^*_{n,\bolds{\tau}^{(n)}}\bigr] \rightarrow_p 0, \qquad\mbox{as }
n\rightarrow\infty,
\end{equation}
where $\rho_M^2$ is the Mallows' metric
described in Section \ref{sec:stationary}. That is, for inferential
purposes, $T[\mathbf{X}^*(\hat{\bolds{\tau}})]$ has approximately
the same
distribution as $T[\mathbf{X}^*(\bolds{\tau})]$. Then, since we have
shown in
Section \ref{secconsistrue} that
\[
\rho_M^2\bigl[F^*_{n,\bolds{\tau}^{(n)}},\Phi(\Sigma_n)\bigr] \rightarrow_p 0,
\]
where
$\Phi(\Sigma_n)$ is the Gaussian distribution with mean 0 and
variance $\Sigma_n$, (\ref{conv1}) implies that
\[
\sqrt{L_n}\Sigma^{-1}_n\bigl\{T\bigl[\mathbf{X}^*\bigl(\hat{\bolds{\tau
}}^{(n)}\bigr)\bigr] -
E^*T[\mathbf{X}^*(\mathbf{t})]\bigr\} \rightarrow N(0,I).
\]

Let $\hat{n}_{i}= \hat{\tau}^{(n)}_{i+1}-\hat{\tau}^{(n)}_i$. We
now state conditions on the
estimated segmentation which guarantee (\ref{conv1}):
\begin{enumerate}[A10.]
\item[A9.] $\hat{U}_n/n \rightarrow0$,\vspace*{1pt}
\item[A10.] $n^{-1} \sum_{i: \hat{n}_i \leq k} \hat{n}_i
\rightarrow0$
for all $k<\infty$,\vspace*{1pt}
\item[A11.] $L_nn^{-1}\sum_{i=1}^{U_n}\min_{1\leq j\leq
\hat{U}_n}|\tau_i - \hat{\tau}_j|\rightarrow_p 0.$
\end{enumerate}
Assumptions A9 and A10 for $\hat{\bolds{\tau}}^{(n)}$ mirror
assumptions A3 and A4
for $\bolds{\tau}^{(n)}$. Assumption A11 is a consistency criterion:
As the data
set grows, the total discrepancy in the estimation of $\bolds{\tau
}^{(n)}$ by
$\hat{\bolds{\tau}}^{(n)}$ must be small.

\begin{theorem} \label{couplethm} Under assumptions \textup{A1--A11},
\textup{(\ref{conv1})} holds.
\end{theorem}

The proof is given in the supplemental article [Bickel et al. (\citeyear{2010bickel-aoas})].
There are trivial extensions of this theorem to smooth functions of
vector means, which are, in fact, needed but simply cloud the
exposition.

Theorem \ref{couplethm} implies that confidence intervals based on
subsamples
\[
\bigl\{\mathbf{X}^{*,j}\bigl(\hat{\bolds{\tau}}^{(n)}\bigr)\dvtx j=1,\dots,B\bigr\}
\]
constructed
by Algorithm \ref{algsubsample} conditional on $\hat{\bolds{\tau
}}^{(n)}$ are
consistent, as long as $\hat{\bolds{\tau}}^{(n)}$ satisfies A9--A11.
Here is the
formal statement of this fact in the one-dimensional case, where
$\hat{\sigma}^2_n$ replaces $\hat{\Sigma}_n$ and $\mathbf{g}$ is the
identity.

\begin{cor} Under assumptions \textup{A1--A11}:
\begin{enumerate}
\item Let $\hat{\sigma}_n^2$ be the block subsampling estimate of
variance defined in \textup{(\ref{subsampvar})}, then
\[
P\bigl(\bar{X}-z_{1-\alpha/2}\hat{\sigma}_n/\sqrt{n}<\mu<\bar
{X}+z_{1-\alpha/2}\hat{\sigma}_n/\sqrt{n}\bigr)
\rightarrow_p 1-\alpha.
\]
\item Confidence intervals estimated by
Efron's percentile method are consistent. That is,
\[
P\bigl(\bar{X}^{*}_{([n\alpha/2])}<\mu<
\bar{X}^{*}_{([n(1-\alpha/2)])}\bigr) \rightarrow_p 1-\alpha.
\]
\end{enumerate}
\end{cor}

%s4.3 ###
\subsection{Segmentation methods}\label{secseg}

The objective of the segmentation step is to divide the original
data sequence $X_1,\dots,X_n$ into approximately homogeneous
\mbox{regions}
so that the variance estimated in Algorithm \ref{algsubsample}
approximates the true variance of $T_n$. A segmentation into regions
of constant mean is sufficient for guaranteeing that Algorithm
\ref{algsubsample} gives consistent variance estimates. Therefore,
we focus here on the segmentation of $X$ into regions of constant
mean.

In our simulation and data analysis, we use the dyadic segmentation
approach, which we motivate and describe here using the simple case
where $g$ is the identity function. First consider the simple case where
$X_1,\dots,X_n$ are independent with variance 1. In testing the null
hypothesis
\[
H_0\dvtx E[X_i] = \mu,
\]
versus the alternative $H_A$ that
there exists $1<\tau<n$ such that $E[X_i]=\mu_1$ for $i<\tau$ and
$E[X_i] = \mu_2 \neq\mu_1$ for $i\geq\tau$, one can show that the
following is the generalized likelihood ratio test:
\[
\mbox{Reject
$H_0$ if } \max_{1<j<n} n M(j) > c,
\]
where
%
%e4.7 ###
\begin{equation}
\label{M} M(j) = \frac{j}{n}(\bar{X}_{1:j}-\bar
{X}_{1:n})^2 +
\frac{n-j}{n}(\bar{X}_{j+1:n}-\bar{X}_{1:n})^2.
\end{equation}
The maximum likelihood estimate of the changepoint parameter $\tau$
is the value that maximizes $M(j)$.

Our proof of Theorem \ref{couplethm} in the supplemental article
[Bickel et al. (\citeyear{2010bickel-aoas})] shows that, in
the case where there is one true change in mean at $\tau$, the
increase in the variance estimated by block subsampling with block
length $L$, given no segmentation [i.e., $\mathbf{t}^{(n)}= \{0,n\}$]
over the
variance estimated by Algorithm \ref{algsubsample} conditioned on a
change-point at $\tau$, is $LM(\tau)+o_p(1)$. Subsampling
conditioned on any segmentation $t \neq\tau$ would inflate the
variance estimate. Hence, segmenting at $\hat{\tau}=\operatorname{arg\,max}_j
M(j)$ is optimal in the sense that $\hat{\tau}$ is the changepoint
estimate that minimizes the asymptotic error of the block subsampling
variance estimate. This fact does not require the assumption of
independence observations, and is true for any second order stationary sequence.
Thus, if we knew that there were only one changepoint, and if the
goal of the segmentation is to obtain the best stratified variance
estimate, then the best place to segment is $\hat{\tau}$. The block
subsampling
variance estimate, given the segmentation $\{0,t,n\}$, would be
\begin{eqnarray} \label{G1} V(t) &=&
\biggl(\frac{t}{n^2}\biggr)\sum_{i=1}^{t-tL/n}(\bar
{X}_{i:i+tL/n}-\bar{X}_{1:t})^2\nonumber
\\[-8pt]\\[-8pt]
&&{}+\biggl(\frac{n-t}{n^2}\biggr)\sum_{i=t+1}^{n-(n-t)L/n}\bigl(\bar
{X}_{i:i+(n-t)L/n}-\bar{X}_{t+1:n}\bigr)^2.\nonumber
\end{eqnarray}

The dyadic segmentation procedure recursively applies the above
logic, as described below.

\begin{alg} \label{algseg} Fix minimum
region length $0<L_s<n$ and threshold \mbox{$b>0$}. Initialize $\mathbf{t}= \{t_0=0,
t_1=n\}$. Repeat:
\begin{enumerate}
\item For $i=1,\dots,|\mathbf{t}|-1$, let $M^{(i)}(j)$ and
$V^{(i)}(j)$ be
respectively the
processes (\ref{M}) and (\ref{G1}) computed on the subsequence
$X_{t_{i-1}+1},\dots,X_{t_i}$.
If $t_{i}-t_{i-1}>2L_s$, then let $t_i' = \operatorname{arg\,max}
_{t_{i-1}+L_s<j<t_i-L_s} M^{(i)}(j)$, $B_i = M^{(i)}(t_i')$ and $V_i =
V^{(i)}(t_i')$.
Otherwise,
let $B_i = 0$, $V_i = \infty$.

\item Let $\lambda_i = L(t_i-t_{i-1})/n$, and
\[
J_i = 1\biggl(\frac{(t_i-t_{i-1})B_i}{\sqrt{V_i\hat{\lambda}_i}} >
b\biggr).
\]
If $\prod_{i} J_i = 0$, stop and return $\mathbf{t}$.
\item Let $i^*=\operatorname{arg\,max}_i B_i$, and $t^{\mathrm{new}} = t_{i^*}'$.
\item Let $\mathbf{t}= \mathbf{t}\cup t^{\mathrm{new}}$,
reordered so that
$t_i$ is monotonically increasing in $i$.
\end{enumerate}
\end{alg}

Each step of the recursion in Algorithm
\ref{algseg} proceeds as follows: In step 1, $M^{(i)}(j)$, the
generalized likelihood ratio process, and $V^{(i)}(j)$, the block subsampling
variance process, are computed for each segment $[t_{i-1}+1,t_i]$
of the current segmentation. For each segment $i$, $B_i$ is the maximum
squared difference
in mean for segment $i$, $t_i'$ is the changepoint estimate that
achieves this maximum, and $\hat{\lambda}_i V_i$ is the estimate of variance
given a changepoint at $t_i'$. In computing $B_i$ and $V_i$ we do
not allow break points that create a region with length less than
$L_s$. In step 2, we normalize the statistic $(t_i-t_{i-1}) B_i$ by
the estimated standard deviation $\sqrt{\hat{\lambda}_i V_i}$. If this
normalized statistic is below the threshold $b$ for every subsegment,
then the recursion stops and returns the current segmentation.
Otherwise, in step 3, the optimal changepoint over all regions
$t^{(\mathrm{new})}$ is
chosen to be the cut that maximizes the decrease in error of the
block subsampling variance estimate. In step 4, this new
changepoint $t^{(\mathrm{new})}$ is added to the current segmentation $\mathbf{t}$.

The computation of $V_i$ in step 2 requires an appropriate choice
$L=L_b$ of the block subsampling sample size. If the correct
segmentation is known, then the choice of $L_b$ is easier, as described
in Section~\ref{sec4.6}. When the segmentation is not known, but a ball park
value of $L_b$ is available, then a segmentation can be computed using
the ballpark value. The segmentation can then be used to obtain a
better choice of $L_b$. If a ball
park value of $L_b$ is not available, then the normalization by
$V_i$ can be omitted, in which case the parameter $b$ in step 3
should be set to 0. This would be equivalent to stopping the
segmentation only when the next optimal cut will violate the minimum
region length $L_s$. In the examples of Section~\ref{sec5.1} we set $b=0$,
thus decoupling the choice of $L_s$ from that of $L_b$.

Two more parameters required by Algorithm \ref{algseg} are $L_s$ and
$b$. The choice for $L_s$ is discussed in Section \ref{secchoosel}.
The choice of $b$ can be guided by the fact that, under the null
hypothesis, if $L$ were chosen appropriately, then
$(t_i-t_{i-1})M^{(i)}(j)/[V^{(i)}(j)\hat{\lambda}_i]^{1/2}$ is a
pivot with
approximate distribution $\chi^2_1$. Asymptotic approximations for
the family-wise error rate have been derived in the case of
independent sequences [James et al. (\citeyear{1987james})]. In the case of dependent
sequences a Bonferroni adjustment can be applied to adjust  for
multiple testing. We also used the formulas given in James et al.
(\citeyear{1987james}) to get a crude cutoff, which seems to work in practice.

Algorithm \ref{algseg} belongs to the class of dyadic segmentation
algorithms for detection of changepoints, the consistency of which
were studied by Vostrikova (\citeyear{1981vostrikova}). These algorithms are greedy
procedures that avoid the search over all possible segmentations. They
have been applied
successfully to various settings in biology, including segmentation
of GC content [Li et al. (\citeyear{2002li})] and the analysis of DNA copy number
data [Olshen et al. (\citeyear{2004olshen})].

The consistency of Algorithm \ref{algsubsample} requires conditions
A9--A11 to be satisfied by the estimated segmentation. The key
condition is A11 which defines a consistency criterion on the
segmentation. Consistency of dyadic segmentation has been proved in
Vostrikova (\citeyear{1981vostrikova}) for sequences that satisfy the following
conditions:
\begin{enumerate}
\item Let $\epsilon_t = X_t-E[X_t]$, then $\| \epsilon_t\|^2$ is
a submartingale and $E\|\epsilon_t\|^2<ct^{\beta}$, $c>0$,
$\beta<2$.
\item The number of regions is fixed and the region sizes are of order
$n$, that is,
\[
\tau_n = (nr_1,\dots,nr_U), \qquad 0<r_1<\cdots<r_U.
\]
\end{enumerate}
It is easy to verify that condition 1 is satisfied
by the piecewise stationary model due to the mixing condition A2.
Condition 2 is more stringent than our assumptions A3 and A4, under
which $U_n$ is allowed to increase with $n$. The consistency of
dyadic segmentation for the case of $U_n \rightarrow\infty$ has
been explored in Venkatraman (\citeyear{1992ven}), who gave asymptotic conditions
on the rejection threshold and on the sizes of the regions to ensure
consistency under the assumption of an independent Gaussian
sequence. However, these conditions are hard to verify in practice,
and for many applications in genomics the more
stringent condition of Vostrikova (\citeyear{1981vostrikova}) is sufficient. Previous
studies on segmenting the genome based on features such as the GC
content [Fu and Curnow (\citeyear{1990fu}); Li et al. (\citeyear{2002li})] have used this
finite regions assumption to achieve reasonable results.

The dyadic segmentation procedure uses information from the entire
sequence to call the first change, and then recursively uses all of
the information from each subsegment to call each successive change
in that segment. An alternative is to use pseudo-sequential
procedures, which are sequential (online) schemes that have been
adapted for changepoint detection when the entire sequence of a
fixed length is completely observed. The basic idea of this class of
methods is to do a directional scan starting at one end of the
sequence. Every time a changepoint is called, the observations
prior to the changepoint are ignored and the process starts over to
look for the next change after the previously detected changepoint.
Specifically, let $\hat{\tau}_0=0$ and, given $\hat{\tau}_1,\dots
,\hat{\tau}_k$,
\[
\hat{\tau}_{k+1} = \inf\{l>\hat{\tau}_k\dvtx
S(X_{\hat{\tau}_k},X_{\hat{\tau}_{k+1}},\dots,X_{\hat{\tau}_l}) >
b\},
\]
where $S$ is a suitably defined changepoint statistic and
$b$ is a pre-chosen boundary. The estimates from pseudo-sequential
schemes depend on the direction in which the data is scanned. Thus,
while they may be suitable for, say, timeseries data, they may not
be natural for segmentation of genomic data, which in most cases do
not have an obvious directionality. The consistency of
pseudo-sequential procedures has been studied by Venkatraman (\citeyear{1992ven}),
who gave conditions on $b=b_n$ and $\hat{\bolds{\tau}}^{(n)}$ for
consistency of
$\hat{\bolds{\tau}}^{(n)}$ under the setting that $X_i$ are
independent Gaussian with
changing means.

%s4.4 ###
\subsection{Testing the null hypothesis of no associations} \label
{hypotesting}

Here we extend the results in Section \ref{SegBlock} to testing the null
hypothesis of no association using nonlinear statistics. As we
discussed in Section~\ref{secexamples}, the inference problem typically posed in
high-throughput genomics is that of association of two features. In
terms of our framework we have two 0--1 processes
$\{I_k\}_{k=1,\ldots,n}$ and $\{J_k\}_{k=1,\ldots,n}$ both defined on a
segment of length $n$ of the genome. We assume that the joint
process $\{I_k,J_k\}$ is piecewise stationary and mixing and want to
test the hypothesis that the two point processes
$\{I_k\}_{k=1,\ldots,n}$ and $\{J_k\}_{k=1,\ldots,n}$ are independent. We
have studied two fairly natural test statistics in ENCODE, the
``percent basepair overlap,''
\[
B_n=\frac{\sum_{k=1}^{n}I_kJ_k}{\sum_{k=1}^{n}I_k},
\]
and the
``regional overlap,'' $R_n$, which we define in Section~\ref{seclinearstat}, with large
values of these statistics indicating dependence. The
major problem we face in constructing a test is what critical values
$o_{n\alpha},r_{n\alpha}$ we should specify so that
%
%e4.8 ###
\begin{equation} \label{eq:8}
P_{H_0}[B_n\geq o_{n\alpha}] \approx\alpha,
\end{equation}
where $H_0$ is the hypothesis that the vectors $(I_1,\ldots,I_n)^T$ and
$(J_1,\ldots,J_n)^T$ are independent, and the corresponding $r_{n\alpha}$
for $R_n$.

We aim for statistics based on $B_n$, $R_n$ (respectively) which are
asymptotically Gaussian with mean 0 under $H_0$. In general, we have to
be careful
about our definition of independence. If we interpret $H_0$ as we
stated, simply
as independence of the vectors $(I_1,\ldots,I_n)^T$ and $(J_1,\ldots
,J_n)^T$, then
\begin{eqnarray*} E_{H_0}(B_n) \approx\frac{\sum_{i=1}^{U}\sum
_{k=1}{n_i} E_{H_0}(I_{ik})E_{H_0}(J_{ik}) }{\sum_{i=1}^{U}\sum
_{k=1}{n_i} E_{H_0}(I_{ik})},
\end{eqnarray*}
where $I_{ik}$ and $J_{ik}$ refer to the $k$th basepair in the $i$th segment
and, hence, we have
%
%e4.9 ###
\begin{equation}\label{neweq4.8}
 E_{H_0}(B_n) \approx\frac{\sum_{i=1}^{U} \lambda_i
E_{H_0}^{(i)}(I) E_{H_0}^{(i)}(J)}{\sum_{i=1}^{U} \lambda_i E_{H_0}^{(i)}(I)}.
\end{equation}
The natural estimate of this expectation is then
\begin{eqnarray*} \frac{1}{\bar{I}} \sum_{i=1}^U \lambda_i \bar
{I}_i \bar{J}_i,
\end{eqnarray*}
where $\lambda_i \equiv\frac{n_i}{n}$, $\bar{I}_i$ is the average
of $I_{ik}$,
$\bar{J}_i$ is the average of $J_{ik}$, and $\bar{I}$ is the grand
average. We assume the correct segmentation.

We proceed with construction of a test statistic and estimation of
the null distribution. In view of \eqref{neweq4.8}, our test
statistic based on $B_n$ is
%
%e4.10 ###
\begin{equation}\label{neweq4.10}
 T_n^O \equiv
n^{{1/2} } (B_n - \widetilde{J}_n ),
\end{equation}
where
\begin{eqnarray}\label{neweq4.11}
 &&\widetilde{J}_n \equiv\Biggl( \sum^{\hat U}_{i=1} \hat{\lambda}_i
\hat{\bar I}_i
\hat{\bar J}_i \Biggr) \bigg/ \frac{1}{n} \sum^n_{k=1} \hat I_k,\nonumber
\\[-8pt]\\[-8pt]
 &&\eqntext{\qquad{}\mbox{where }\hat{\lambda}_i= \lambda_i(\hat
\mathbf{t}),
\hat{\bar I}_i=n^{-1}_i(\hat\mathbf{t}) \sum^{\hat t_i}_{k= \hat
t_{i-1}+1} I_k\nonumber}
\end{eqnarray}
with $\hat{\bar J}_i$ similarly defined. Here is
the algorithm based on this statistic.

\begin{alg} \label{algtesting}
In order to estimate the null distribution, we do the following:
\begin{enumerate}
\item Pick at random without replacement two starting points, $K_1$ and
$K_2$, of blocks
of length $L$ from $\{1,\ldots,n-L\}$.
\item Let $(I_{{K_1}+1},\ldots,I_{{K_1}+L})^T$ and
$(J_{{K_1}+1},\ldots,J_{{K_1}+L})^T$,
$(I_{{K_2}+1},\ldots,I_{{K_2}+L})^T$ and $(J_{{K_2}+1},\ldots,J_{{K_2}+L})^T$
be the two sets of two feature indicators.
%Consider $B_n$ with $R_n$ being treated analogously.
%
\item Form
\begin{eqnarray*}
\overline{IJ}_{nL}^{*1} &\equiv&\frac{1}{L} \sum^L_{l=1} I_{K_1 +l}
J_{K_2 +l},
\\
\bar{I}_{nL}^{*1} &\equiv&\frac{1}{L} \sum^L_{l=1} I_{K_1 +l},
\\
\overline{IJ}_{nL}^{*2} &\equiv&\frac{1}{L} \sum^L_{l=1} I_{K_2 +l}
J_{K_1 +l}
\end{eqnarray*}
and define $\bar{I}_{nL}^{*2}$, $ \bar{J}_{nL}^{*1} $, $
\bar{J}_{nL}^{*2} $ analogously. Let
\begin{eqnarray*}
F_{nL}^* &\equiv&\frac{1}{2} \biggl( \frac{\overline{IJ}_{nL}^{*1}
}{\bar{I}_{nL}^{*1} }
+ \frac{\overline{IJ}_{nL}^{*2}}{\bar{I}_{nL}^{*2} } \biggr),
 \\
T_{nL}^* &\equiv& F_{nL}^* - \bar{J}_{nL}^*,
\end{eqnarray*}
where
\[
\bar{J}_{nL}^* = \frac{1}{2} (\bar{J}_{nL}^{*1} + \bar{J}_{nL}^{*2}
)
\]
and $\bar{I}_{nL}^*$ is defined analogously. Let $F_{nLb}^*$,
$\overline{IJ}_{nLb}^{*1}$, etc., be obtained by choosing
$(K_{1b},K_{2b})$, $b=1,\ldots,B$, independently as usual.
%$$O_{nL}^*=\frac{\sum_{l=1}^L(I_{k_1+l}J_{k_2+l}+I_{k_2+l}J_{k_1+l})}
%{\sum_{l=1}^L(I_{k_1+l}+I_{k_2+l})}$$
% and let $O_{nL1}^*,\ldots,O_{nLB}^*$ be obtained by choosing
% $(K_{1b},K_{2b})$, $b=1,\ldots,B$ independently as usual.
% Define $\bar{O}_{nL}^*=\frac{1}{B}\sum_{b=1}^B O_{nLb}^*$,
% $\widetilde{O}_{nLb}^*=O_{nLb}^* - \bar{O}_{nL}^*$, $b=1,\ldots,B$
% and write $\widetilde{O}_{nL}^*$ for a single
% $\widetilde{O}_{nLb}^*$.
%
\item We use the following $c_{nL\alpha}$ as a critical value for
$B_n$ at level
$\alpha$,
\[
c_{nL\alpha}=\bar{J}_n+\biggl(\frac{2L}{n}\biggr)^{{1/2}}
T^*_{nL(B(1-\alpha))}  ,
\]
where $T_{nL(1)}^* \leq\cdots \leq
T_{nL(B)}^*$ are the ordered $T_{nLb}^*$ and $[\cdot]$ denotes integer
part and $\bar{J}_n=\frac{1}{n}\sum^n_{k=1} J_k$.
\item If the sequence is piecewise stationary with estimated segments
$j=1,\ldots,\hat{U}_n$ as in Section \ref{secseg}, we draw independently $B$
sets of starting points, $K_{11}^{(j)},\ldots,\break K_{1B}^{(j)}$ and
$K_{21}^{(j)},\ldots,K_{2B}^{(j)}$, of blocks of length
$\hat{\lambda}_jL$ from each segment $i=1,\ldots,j$ when each pair
is drawn at random without replacement. Here $\sum_{i=1}^U
\hat{\lambda}_i =1 $ and $\hat{\lambda}_i$ is proportional
to the length of estimated segment $i$. Then piece $T^*_{nLb}$
together as follows. Let
\begin{eqnarray*}
\overline{IJ}_{nLb}^{*1i} &=& \frac{1}{L\hat{\lambda}_i}
\sum_{l=1}^{ \hat{\lambda}_i}
I_{iK_{1b}+l} J_{iK_{2b}+l},
\\
\bar{I}_{nLb}^{*1i} &=&\frac{1}{L\hat{\lambda}_i}
\sum_{l=1}^L I_{iK_{1b}+l},
\\
&&\quad\mbox{etc.},
 \\
\bar{F}_{nLb}^* &=& \sum^{\hat U}_{i=1} \hat{\lambda}_i \biggl(
\frac{\overline{IJ}_{nLb}^{*1i} } {\bar{I}_{nLb}^{*1i} } +
\frac{\overline{IJ}_{nLb}^{*2i} } {\bar{I}_{nLb}^{*2i} } \biggr)  .
\end{eqnarray*}
Then,
\[
T^*_{nLb} = F^*_{nLb} - \widetilde{J}^*_{nLb}  ,
\]
where
\[
\widetilde{J}^*_{nLb} = \frac{\sum_{i=1}^{\hat U}(\bar
I^{*i}_{nLb})(\bar J^{*i}_{nLb})\hat\lambda_i}{\sum_{i=1}^{\hat
U}(\bar I^{*i}_{nLb})\hat\lambda_i}
\]
with $\bar I^{*i}_{nLb} =
\bar I^{*1i}_{nLb} + \bar I^{*2i}_{nLb} $. The critical value is
\[
\widetilde{J}_n+ \biggl( \frac{2L}{n}\biggr)^{{1/2}}
T^*_{nL(B(1-\alpha))}  ,
\]
as before.
\end{enumerate}
\end{alg}

We can apply this principle more generally to statistics which are
functions of sums of products of $I$'s and $J$'s evaluated at the
same positions.

The proof of the following theorem is given in the supplemental article
[Bickel et al. (\citeyear{2010bickel-aoas})].

\begin{theorem}
\label{theorem4.9}
If $\mathcal L_0$, $P_0$ denote distributions under the hypothesis of
independence and \textup{A1--A11} hold, then
\begin{enumerate}
\item[1.] $\mathcal L_0 (T_n^O) \Longrightarrow\mathcal N(0,\sigma^2_0)$
\item[2.] With probability tending to 1,
\[
\mathcal L_0^*(T^{O*}_{n,L}) \Longrightarrow\mathcal N(0,\sigma^2_0).
\]
\item[3.] $P_0 [ T_n^O \geq( \frac{2L}{n} )^{{1/2}}
\hat q_{1-\alpha}^0] \to\alpha$ where $\hat q_{1-\alpha}^0$ is
the $[ (1-\alpha)B]$th of $T^{O*}_{nLb}$, $1\leq b \leq B$.
\end{enumerate}
\end{theorem}

%$T_n^R$, $T^{O*}_{n,L}$ replaced by $T^{R*}_{n,L}$ etc. and the
%vector having added the components ($ \frac{1}{n} \sum^n_{i=2} I_i
%I_{i-1}J_i J_{i-1}$, $\frac{1}{n} \sum^n_{i=2} I_iI_{i-1}$,
%$\frac{1}{n} \sum^n_{i=2} J_i J_{i-1}$).

In practice, this definition of independence makes our statistic in effect
reflect \emph{conditional independence} of $I_k$ and $J_k$ given the
segment to which
the $k$th base belongs. This can be unsatisfactory in practice, for
instance, when the features
are concentrated in small segments such that large, sparse segments
swamp the inference.

We define \emph{independence irrespective of segment identity} as
saying that the average over all permutations of the segments of the
joint distribution of the point process features are independent.
Formally, if $(P_1,\ldots,P_U)$, $(Q_1,\ldots,Q_U)$ denote the marginal
distributions of $\{ \{ I_{ik}\dvtx k=1,\ldots,n_i \} \dvtx i=1,\ldots,U \}$ and
$\{ \{ J_{ik}\dvtx k=1,\ldots,n_i \} \dvtx i=1,\ldots,U \}$, and $(R_1,\ldots,R_n)$
correspond to the joint\vspace*{1pt} distribution of $\{(I_{ik},J_{ik})\dvtx 1 \le k
\le n \}$, then let $(\bar{P}_1,\ldots,\bar{P}_U) = \frac{1}{U!} \sum
{(P_{\pi1},\ldots,P_{\pi U})}$ where $\pi$ ranges over all
permutations of $1,\ldots,U$. Define $(\hat{Q}_1,\ldots,\hat{Q}_U)$ and
$(\hat{R}_1,\ldots,\hat{R}_U)$ similarly. Then, our hypothesis is
%
%e4.11 ###
\begin{equation} H_1 \dvtx  \hat{R} = \hat{P}
\times\hat{Q}.
\end{equation}
This is simply saying
that independence is not conditional on relative genomic position of
segments.

It is easy to see that we should now define
%
%e4.12 ###
\begin{equation} T_n^{\widetilde{O}} = n^{{1/2}}(B_n - \hat
{J}_n),
\end{equation}
where $\hat{J}_n = \frac{1}{n}\sum_{i=1}^{n}J_i$.

The reason for this is that
%
%e4.13 ###
\begin{equation} E_{\hat{R}}(B_n) \approx\frac
{E_{\hat{R}}(\frac{1}{n}\sum_{i=1}^{n} I_i
J_j)}{E_{\hat{R}}(\hat{I})}.
\end{equation}
Under $H_1$,
\begin{eqnarray*}
E_{\hat{R}}\Biggl(\frac{1}{n}\sum_{i=1}^{n} I_i J_j\Biggr) =
E_{\hat{P}}(\hat{I})E_{\hat{Q}}(\hat{J})
\end{eqnarray*}
and
\begin{eqnarray*} E_{\hat{R}}(I) = E_{\hat
{P}}(\hat{I}),
\end{eqnarray*}
so that the statistic simplifies to the $U=1$ form, as above.

It is clear that the conclusion of (\ref{eq:8}) continues to hold when
applied to $T_n^{\widetilde{O}}$. Note that the form of the
bootstrap is unchanged, since $T_n^{\widetilde{O}}$ is invariant
under permutation of the segments.

We now turn to $R_n$ as defined in Section~\ref{seclinearstat}. We assume that ${V_i \dvtx
i=1,\ldots,K}$ are strongly mixing and stationary. If we assume $H_0$,
we have no closed form for $E_{H_0}(\frac{1}{W}\sum_{k=1}^K V_k)$ by
which to center $R_n$. To estimate this quantity, we apply a version
of the double bootstrap [Beran (\citeyear{1988beran}); Hall (\citeyear{1992hall}); Letson and
McCullough (\citeyear{1998letson})].

Consider $\frac{1}{n} \sum_{k=1}^K V_k$ under $H_1$. We draw $B_1$
pairs of large blocks of length $mL$, and we compute the \% false
region overlap, call it $R_b^*, b=1,\ldots,B$, in each pair of
``large''
blocks, where $mL$ is still negligible compared to segment size, but
$m \rightarrow\infty$. Define
%
%e4.14 ###
\begin{equation}
\hat{E}_{H_1} (R_n) = \frac{1}{2B} \sum_{b=1}^{B_1} R_b^*
\end{equation}
and
%
%e4.15 ###
\begin{equation} \widetilde{T}_n^{(R)} =
n^{{1/2}}\bigl(R_n - \hat{E}_{H_1} (R_n)\bigr).
\end{equation}

Note that we again want to consider independence irrespective of segment
identity, so that $R^*_b$ above are computed without any segmentation beyond
the natural segmentation, for example, chromosomes.
Now compute the empirical distribution of $\widetilde{T}_n^{(R)}$ using
the size $L$ segmented block subsampling and proceed as usual. We can
define $\widetilde{T}_n^{(R)}$ corresponding to $H_0$ in the same
way, though we now have to cut up our $mL$ blocks in proportion to
segment sizes to center. We do not pursue this since the $H_1$
hypothesis gives stable results while $H_0$ does not.

We have not proved a result justifying the use of the double bootstrap
in this way, but simulations suggest that it behaves as expected; see
Section~\ref{secsimIIb}.

%s4.5 ###
\subsection{Choice of segment size $L_s$} \label{secchoosel}
 Two tuning parameters appear in our procedure in addition to $b$
appearing in the segmentation scheme. $L_s$ is the smallest allowed
size of a ``stationary'' piece after
segmentation. It essentially determines the scale of the
segmentation, which we view as an application context dependent
quantity that users need to control. The reason is that stationarity
is a matter of scale. To put it concretely, consider the situations
where $I_k$, $k=1,\dots,n$, are simply the base pair nucleotides
$A,C,G,T$ and consider the scale of a large gene of length $n$.
Then, it seems natural that the exons and introns correspond to
consecutive stationary regimes. However, suppose we now move our
scale to a gene rich genomic region of length $N$. Now, it is the
genes themselves and the intergenic regions which correspond to an
at least initial segmentation.

This dependence of segmentation on scale has a natural intuitive
consequence. Consider a statistic such as base pair overlap of two
features. As one increases the region size $n$ in which one wishes
to declare significant overlap, the standard deviation of the
statistic, which is $O(n^{-1/2})$, decreases, and $p$-values decrease.
However, if, as one would expect, the region over which $n$
increases becomes homogeneous on a larger scale, coarser
segmentation would then be called for. This, as we have noted,
necessarily increases the standard deviation of the statistic, and
from that point of view significance becomes more difficult to
achieve.

Put another way, it is not impossible to think of the whole genome
itself as being stationary on a large scale, but that we can
hierarchically segment the genome in many ways so that each large
subsegment is stationary, but the segments are not identically distributed,
even where they are of equal length. For instance, a natural initial
segmentation is to chromosomes.

Finally, we argue in mathematical terms going the other way from
inhomogeneity to homogeneity. Start with a sequence of independent
(say) Bernoulli variables $X_1,X_2,\dots,X_n$, with $X_k$ being
$\operatorname{Bernoulli}(p_k)$. If the $p_k$ are arbitrary, the only segmentation
we can perform is the useless trivial one, where each $X_k$ is its
own segment. But, now  we tell ourselves that $p_k$,
$1\leq k\leq n/2$, are drawn i.i.d. from $U(0,1/2)$ and for
$n/2+1\leq k \leq n$ from $U(1/2,1)$, we suddenly just have two
segments to consider.

Thus, $L_s$ in our view needs to be treated as the smallest scale on
which homogeneity is expected. Note that these considerations are
not limited to testing. They also govern confidence intervals, as
discussed in Section \ref{secmainresult}.

%s4.6 ###
\subsection{Choice of $L_b$, the subsample size}\label{sec4.6}
We believe that the best way to choose $L_b$, after segmentation
has been estimated, is so that the resulting subsampling distribution
of the statistics is
as stable as possible and $L_s$ is large but $\ll n$. We also formally
consider Gaussianity of the distribution and, if possible, maximizing
that feature as well.
This does not necessarily mean segment more---since A10 and A11 may
then fail. We advocate
but do not analyze further the following proposal put forward in
$m$-out-of-$n$ subsampling by Bickel, G\"{o}tze and van Zwet (\citeyear{1997bickel})
and analyzed in detail by G\"{o}tze and Rackauskas (\citeyear{2001gotze}) and Bickel
and Sakov (\citeyear{2008bickel}):

\begin{enumerate}
\item Let $\bar{X}_n^*(L)$ be the statistic computed from the sample
drawn with blocks of length $L$.
Compute the block subsampling distribution $\mathcal{L}_{L_v}$ for the
statistic
\[
\sqrt{L_v}\bigl(\bar{X}^{*}_{n}(L_v) - \bar{X}_n\bigr)
\]
and $L_v = \rho^vn$, where $\rho<1$ and $v=1,2,\dots,V$. Note that
these $L_v$ provide candidate choices of the subsample size $L_b$.
\item Compute a ``distance'' $d^*(v)$ between $\mathcal{L}_{L_v}$ and
$\mathcal{L}
_{L_{v-1}}$.
\item Choose $L_b = L_{v^*_0}$, where $v^*_0=\operatorname{arg\,min} d^*(v)$.
\end{enumerate}

In practice, we use for $d^*(v)$ the pseudometric
\[
\Biggl|\sqrt{ \frac{L_{v-1}}{L_v}}  \mathit{IQR}(\mathcal{L}_{L_v}) -
\mathit{IQR}(\mathcal{L}_{L_{v-1}})\Biggr|,
\]
where $\mathit{IQR}(\mathcal{L})$ is the interquartile range of $\mathcal{L}$.

In continuing work with G\"{o}tze and van Zwet, we are in the process of
trying to show that, under mild conditions, as $n\rightarrow\infty$
we have $L_b \rightarrow\infty$, $L_b/n \rightarrow0$. More
significantly, we expect that in a fashion analogous to G\"{o}tze and
Rackauskas (\citeyear{2001gotze}) and
Bickel and Sakov (\citeyear{2008bickel}), under restrictive conditions and for suitable
choice of distance, $L_b$ yields an estimate which is as good as
possible in the following sense: If $\mathcal{L}_n$ is the actual distribution
of $\sqrt{n}(\bar{X}_n-\mu)$, $d(m)$ is the distance between
$\mathcal{L}_n$ and $\mathcal{L}_{L_v}$, and $v_0 = \operatorname{arg\,min}_v d(v)$, then
\[
\frac{d(v^*_0)}{d(v_0)} \rightarrow_p c.
\]
Thus, $L_{v_0^*} = \rho^{v^*_0}n$ yields performance of the same
order as $\rho^{v_0}n$.

%s5 ###
\section{Simulation and data studies} \label{secsim}

%s5.1 ###
\subsection{Simulation study I}\label{sec5.1} In this section we perform a
simple simulation study to demonstrate the power of our
block-subsampling method in the situation where features are
naturally clustered. We simulated a binary sequence $x_1,\ldots,x_n$
with $n=10{,}000$ by the following Markovian model:
\begin{eqnarray}\label{eq:s1}
P(x_1=1)&=&\frac{p_0}{2},\nonumber
\\[-8pt]\\[-8pt]
P(x_k=1) &=&
\frac{p_0}{2}+(1-p_0)\frac{\sum_{j=k-w}^{k-1}x_j}{w} \qquad\mbox{for }
k=2,\ldots,n,\nonumber
\end{eqnarray}
 where $w$ is the
order of the Markov model or, intuitively, the size of the dependency
window, and $p_0$ indicates the level of dependency. Smaller
$p_0$ gives stronger dependence between neighboring
positions. We define the following two types of features at
position $k$ in the sequence:
\begin{itemize}
\item Feature I: the occurrence of sequence 11,100 starting at
position $k$.
\item Feature II: the occurrence of more than six 1's in the next 10 consecutive
positions including the current position $k$.
\end{itemize}

From model (\ref{eq:s1}), the feature II will occur in clusters in
the sequence. The overlap between the two types of features can be
measured by the statistic
\[
S=\frac{\sum_{k=1}^n I_kJ_k}{\sum_{k=1}^n I_k}
\]
with $I_k$, $J_k$
being binary and indicating the occurrences of sites of types I and
II, respectively.

\begin{figure}

\includegraphics{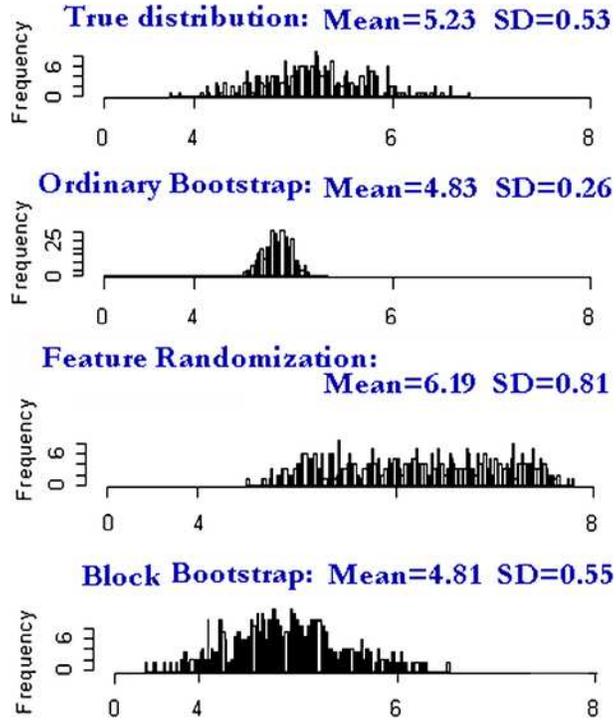}

\caption{Comparison of different subsampling schemes.}\label{fig:sim1:1}
\end{figure}

Figure \ref{fig:sim1:1} shows the distribution of $S$ estimated through different
ways:
\begin{itemize}
\item The true distribution is the empirical distribution of estimated
$S$ from
10,000
random sequences generated under model (\ref{eq:s1}).
\item The Ordinary Bootstrap distribution is derived by performing a
base-by-base uniform sampling
of the sequence $x_1,\ldots,x_n$ to construct 10,000 sequences of length $n$.
\item The Feature Randomization distribution is derived by keeping
features of type I fixed and
randomizing uniformly the start positions of the features of type II
to construct 10,000 sequences of length $n$.
\item The block subsampling distribution is derived by drawing
independent samples of blocks
of length $L=40$ and stringing the blocks together to construct
10,000 sequences of length $n$.
\end{itemize}

%f1 ###

From Figure \ref{fig:sim1:1}, we see that block subsampling produces more reliable
estimates of the
variance of $S$ compared to the naive methods: ordinary
bootstrapping and feature randomization. Both naive methods ignore
the dependence between positions and thus fail to take into
account the natural clumps of the feature II. This is the key reason
for the poor performance of the two naive methods.

%s5.2 ###
\subsection{Simulation study IIa}\label{sec5.2}

\begin{figure}[t]

\includegraphics{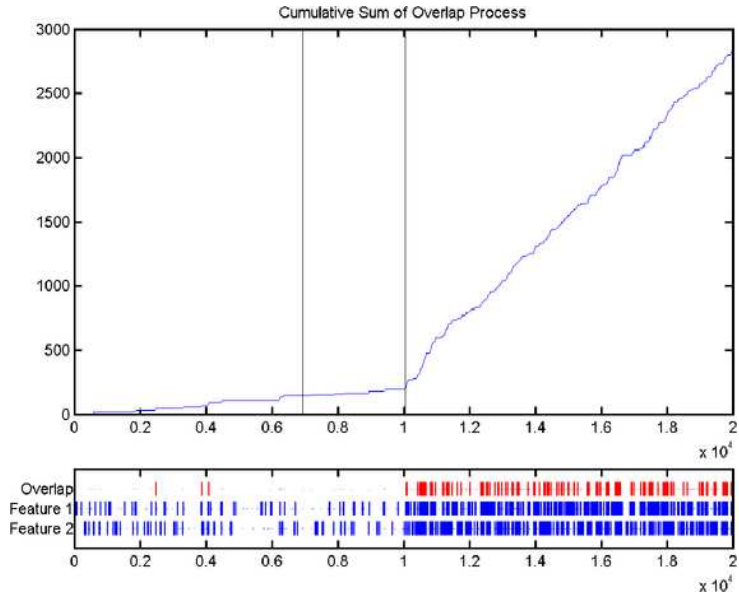}

\caption{Example of one instance from simulation model 2. Top plot
shows cumulative sum and estimated segmentation.}\label{fig:sim2:1}
\end{figure}

\begin{figure}[t]

\includegraphics{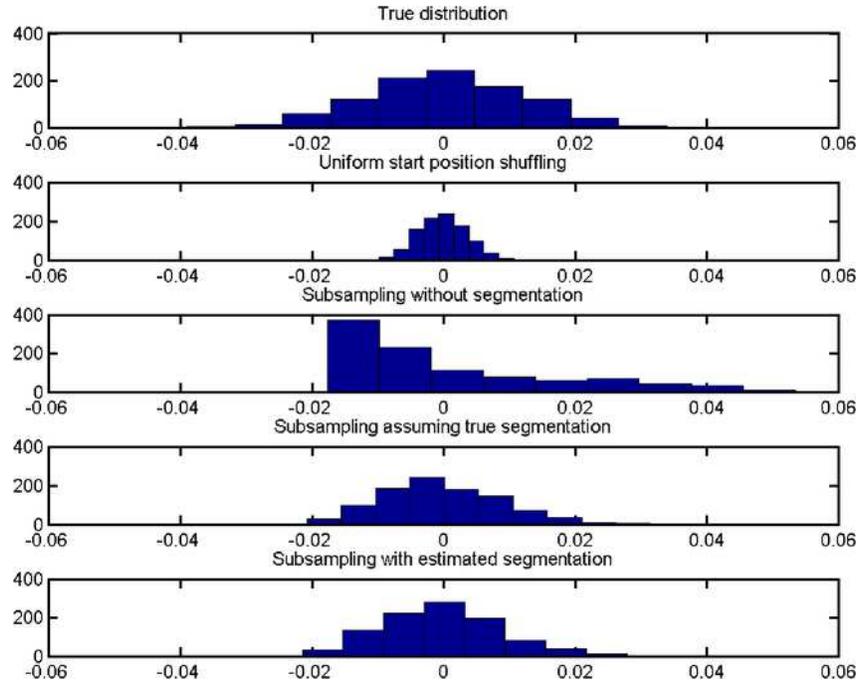}

\caption{Comparison of different subsampling schemes.}\label{fig:sim2:3}
\end{figure}

Our second simulation study examines the case where the sequence is
generated from a piecewise stationary model where there is more than
one homogeneous region. As before, we consider the problem of
estimating the percentage of base pair overlap between two features,
and compare the performance of four strategies:
\begin{enumerate}
\item feature randomization,
\item naive block subsampling from unsegmented sequence,
\item block subsampling from sequence segmented using the true
changepoints, and
\item block subsampling from sequence segmented using the changepoints
estimated by the dyadic segmentation method we described in Section~\ref{secseg}.
\end{enumerate}
In our simulation model, we generate $X_t$, $Y_t$ independently from
a Neyman--Scott process characterized as follows:
\begin{enumerate}
\item Cluster centers occur along the sequence according to a Poisson
process of rate $\lambda_i$ in region $i$.
\item The number of features in each cluster follows Poisson
distribution with mean~$\alpha$.
\item The start of features are located at a geometric distance (mean
$\mu$) from the cluster center.
\item The features are generated with length that is geometric with
mean $\beta$.
\item Overlap between features generated using steps 1--4 are ignored.
\end{enumerate}
For simplicity, we let there be only 2 homogeneous
regions, each of length $T=10{,}000$. Consider the setting where the
parameters for the two regions have the following values:
$(\lambda_1,\alpha_1,\mu_1,\beta_1)=(0.01,10,10,5)$ and
$(\lambda_2,\alpha_2,\mu_2,\beta_2)=(0.02,10,10,5)$. Figure
\ref{fig:sim2:1} shows a simulated example, where features $A$ and $B$
are plotted as well as their overlap. Figure \ref{fig:sim2:1} also
shows the cumulative sum and the segmentation. Figure
\ref{fig:sim2:3} shows respectively the histograms of the estimated
distribution of the overlap statistic $\bar{X}^*$ centered and
scaled. It is clear that the feature randomization underestimates
the standard deviation, whereas naive block subsampling without
segmentation gives a mixture distribution with long tails. Strategy
3, which subsamples assuming the true changepoint at $\tau$ is
known, gives the correct distribution as expected. Strategy 4, which
uses the estimated changepoint, reassurringly gives a very similar
distribution to strategy 3. Table \ref{tab2} gives the
standard deviation estimates.

%f2 ###

%f3 ###

%error by
%four sampling strategies in simulation study 2.}

%s5.3 ###
\subsection{Simulation study IIb} \label{secsimIIb}

We utilized the Neyman--Scott process described in simulation study
IIa to study the consistency of the double bootstrap method
described in Section \ref{hypotesting} for estimating the
distribution of $R_n$. We consider the simple case where there is
one homogeneous region.
% Since the statistic $R_n$ generally depends upon smaller counts
%(unless features are point-processes),
We utilized a larger region and a parameterization of the process
that results in more and longer feature instances than we consider
in the study above. The parameters are $T=5$ Mb and
$(\lambda_1,\alpha_1,\mu_1,\beta_1)=(\lambda_2,\alpha_2,\mu
_2,\beta_2)=(0.05,10,100,75)$.
This yields a pair of feature-sets with around 5000 instances,
where each feature-set covers around 17\% of the 5~Mb region. We
simulated 20,000 pairs of feature-sets from this process, and found
that the mean of region-overlap between pairs, $R_n$, was 0.293, and
the standard error was 0.0072. We subsampled 1000 sets of 10,000
draws from this distribution, each of which yielded the mean above
(to 3 significant digits), and the standard errors ranged from
0.0071 to 0.0073, which corresponds almost exactly to the
theoretical 95\% confidence interval for the standard error of the
standard error of a Gaussian with standard deviation 0.0072 after
10,000 draws. Not surprisingly, the distribution of $R_n$ was
Gaussian, as indicated by the Lilliefors and the Shapiro--Wilk test,
which did not reject the hypothesis of Gaussianity at a significance
level 0.05 with the full sample of 20,000 observations.

\begin{figure}[t]

\includegraphics{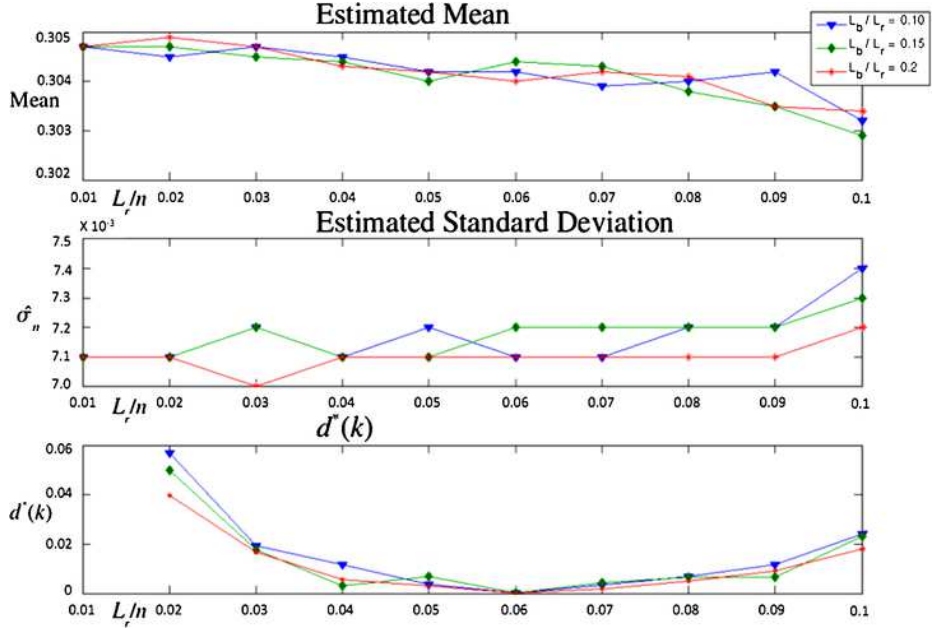}

\caption{Comparison of block subsampling distributions.}\label{table:2}
\end{figure}

In order to test the capacity of segmented subsampling with a version
of the
double block bootstrap to discover this distribution based on only a
single pair of observations, we selected the most extreme pair found
during simulation, for which $R_n$ was 0.321, corresponding to a
$z$-score of 3.87. Since the number of feature instances is itself a
random quantity, the job of block subsampling is particularly
difficult: when $R_n$ is far to the right of expectation, the
feature-sets tend to contain more feature instances than those
closer to the center. The pair we chose was no exception. The
results are given in Figure \ref{table:2}. Hence, it is not surprising
that our subsampling
procedure tends to over-estimate the mean. The
Lilliefors test fails to reject the Gaussianity of any of the
resulting distributions with sample sizes up to 1000 at a
significance level of 0.05, but does reject it for several of the
smaller block-sizes when the sample size is pushed up to 10,000.
The Shapiro--Wilk test, however, detects departures from Gaussianity
for many of the distributions at a significance level of 0.05 for
samples larger than 500. This is because $R_n$ is predicated on
relatively small counts of feature-instance overlaps and, hence, the
distributions tend to have heavy tails.

We note that the global minimum of the Inter-Quantile (IQ) statistic
was found at $L_b/L_r = 0.15$ and $L_r/n = 0.06$. That is, 0.9\% of
the 5~Mb region, or 45~Kb, were included in each block sample. This
block sample size is certainly sufficient to capture multiple
feature-clusters, since the parameterized Neyman--Scott process above
yields an average inter-cluster distance of about 1~Kb.

%t1 ###
\begin{table}[b]
\caption{Estimates of standard error by four sampling
strategies in simulation study IIa}\label{tab2}
\begin{tabular*}{\tablewidth}{@{\extracolsep{4in minus 4in}}lcc@{}}
\hline  & \textbf{Standard error} & \textbf{Fold change from}\\
\textbf{Method}& \textbf{estimate} & \textbf{true value} \\
\hline
True value & 1.2e$-$002 & ---\\
Uniform shuffle & 3.6e$-$003 & 0.3\phantom{0}\\
Subsample, no segmentation & 1.7e$-$002 & 1.4\phantom{0}\\
Subsample, true segmentation & 1.1e$-$002 & 0.91\\
Subsample, estimated segmentation & 1.0e$-$002 & 0.83\\
\hline
\end{tabular*}
\end{table}

%f4 ###

To corroborate our hypothesis that the mean was overestimated
because the feature-sets we chose were more dense than most, we
applied our method with learned parametrization, $L_b/L_r = 0.15$
and $L_r/n = 0.06$, for a pair of feature-sets with $R_n = 0.293$,
the population average. Indeed, the mean was estimated, after
10,000 samples, to be 0.293, and $\hat{\sigma}_n$ was 0.0072.

Although the purpose of this simulation was merely to check the
consistency of our version of the double block bootstrap for data
not unlike actual genomic data, for example, ChIP-seq ``broad-peaks,'' we
decided to also check the performance of feature-start site
shuffling for the same pair of feature-sets used above. In the case
of $B_n$, the basepair overlap statistic, feature start-site
shuffling correctly estimates the mean, but can (in the stationary
case) radically underestimate the standard deviation. The same is
not true in the case of $R_n$. Start-site shuffling is not assured
(under our model) to provide an unbiased estimate of the mean or the
standard deviation. We drew 10,000 samples from the distribution
under shuffling, and found the mean to be 0.337, and the standard
deviation to be 0.0070, which indicates that the pair of
feature-sets under study in fact overlap slightly less than expected
at random ($p \approx0.011$). The fact that this conclusion is
actually in the wrong direction in this relatively easy, stationary
example should make us skeptical of studies that rely upon
start-site shuffling to draw conclusions about statistics that
cannot be defined locally, such as $R_n$.

Our discussion of this simulation and the following real data
examples exhibit the subtleties inherent in our approach. Subtleties
appear whenever inference follows regularization.

%{\small\noindent
% \hline\hline
% & \multicolumn{3}{l|} {$L_b/L_r = 0.1$} & \multicolumn{3}{l|}
%{$L_b/L_r = 0.15$} & \multicolumn{3}{l|} {$L_b/L_r = 0.2$}\\
% \hline
% % Fraction of 5Mb region & \multicolumn{13}{l|} {region overlap
%($R_n$)} \\
% \cline{2-4}
% ($L_r/n$) & $d^*(k)$ & Mean & $\hat{\sigma}_n$ & $d^*(k)$ & Mean & $
% \hline
%0.10 & 0.0240 & 0.3032 & 0.0074 & 0.0231 & 0.3029 & 0.0073 & 0.0181 &
%0.3034 & 0.0072 \\
%0.09 & 0.0116 & 0.3042 & 0.0072 & 0.0064 & 0.3035 & 0.0072 & 0.0092 &
%0.3035 & 0.0071 \\
%0.08 & 0.0070 & 0.3040 & 0.0072 & 0.0067 & 0.3038 & 0.0072 & 0.0049 &
%0.3041 & 0.0071 \\
%0.07 & 0.0032 & 0.3039 & 0.0071 & 0.0042 & 0.3043 & 0.0072 & 0.0017 &
%0.3042 & 0.0071 \\
%0.06 & 0.0003 & 0.3042 & 0.0071 & 0.00004 & 0.3044 & 0.0072 & 0.0002 &
%0.3040 & 0.0071 \\
%0.05 & 0.0036 & 0.3042 & 0.0072 & 0.0070 & 0.3040 & 0.0071 & 0.0031 &
%0.3042 & 0.0071 \\
%0.04 & 0.0115 & 0.3045 & 0.0071 & 0.0031 & 0.3044 & 0.0071 & 0.0057 &
%0.3043 & 0.0071 \\
%0.03 & 0.0194 & 0.3047 & 0.0072 & 0.0178 & 0.3045 & 0.0072 & 0.0169 &
%0.3047 & 0.0070 \\
%0.02 & 0.0571 & 0.3045 & 0.0071 & 0.0498 & 0.3047 & 0.0071 & 0.0396 &
%0.3049 & 0.0071 \\
%0.01 & N/A & 0.3047 & 0.0071 & N/A & 0.3047 & 0.0071 & N/A & 0.3047 &
%0.0071 \\
%{\bf Table 2:} Comparison of block subsampling distributions \\

%s5.4 ###
\subsection{Association of noncoding ENCODE annotations and
constrained sequences}\label{sec5.4}

Here we present a real example of the study of association between
``constrained sequences'' and ``nonexonic annotations'' from the
ENCODE project, limited to the 1.87~Mbp ENCODE Pilot Region ENm001,
also known as the CFTR locus. The constrained sequences are those
highly conserved between human and the 14 mammalian species studied
and sequenced by the ENCODE consortium.
%Evidence for evolutionary constraint in these non-coding annotations
%is of particular
%interest, as, for the most part these annotations do not come with
%proposed biological functions, but rather constitute biochemically
%``active sites" of the DNA molecules in vivo.
Enrichment of evolutionary constraint at the ``nonexonic
annotations'' sites implies that the biochemical assays employed by
the ENCODE consortium are capable of identifying biologically
functional elements. We tested the association of noncoding
annotations and constrained elements using the base pair overlap
statistic $B_n$ in Section~\ref{secseg} using the conditional formulation. We
interpret the lack of association
as, given sequence composition and the distribution of each feature
along the genome as observed, the assignments (by nature) of
features $A$ and  $B$ to individual bases are made independently. We
derive the significance of the observed statistic under this null
hypothesis following the method proposed in Section~\ref{secseg}.

As we discussed, we have several issues to deal with:
\begin{enumerate}[(iii)]
\item[(i)] How do we segment? That is, what statistic(s) do we use for
segmenation?
\item[(ii)] Is segmentation necessary or is the region sufficiently
homogeneous?
\item[(iii)] If we segment, what $L_s$ should we use?
\item[(iv)] Given a segmentation, what $L_b$ is appropriate?
\end{enumerate}

Here are our methods:
\begin{enumerate}[(a)]
\item[(a)] The simplest choice for (i) and the one we followed was to segment
according to both numerator and denominator in $B_n$: intersect
partitions and enforce an $L_s$ bound. Given our theory, this
should ensure homogeneity in the mean of~$B_n$.
\item[(b)] Although strictly speaking (ii) and (iii) can be combined,
we experimented
a bit to also see if the theory of Section~\ref{sec4.1} was borne out in
practice.
\item[(c)] We did not use the $V$ statistic and thus only had to
choose $L_s$.
Again, we experimented with $L_s = 500$~Kb to preserve as much
genomic structure as possible, and $L_s = 200$~Kb to ensure we had
not undersegmented.
\item[(d)] We explored a variety of values of $L_b$, and studied the
consistency
between nearby values under the interquartile statistic (IQ
statistic) discussed in Section~\ref{sec4.6}. We draw conclusions based on
the value of $L_b$ that optimizes local consistency.
\end{enumerate}

To segment the data, we applied the method in Section \ref{secseg}
to both features $A$ and $B$, or in the language of Section~\ref{secsubsampling}, $I$ and $J$,
and then combined the segmentation. In
segmenting each feature, we experimented with minimum segment
lengths $L_s$ of 200 and 500 Kb. Before subsampling, we combined the
segmentations of $A$ and~$B$ by taking a union of the changepoints.
This created regions with length less than $L_s$. However, the total
length of these regions comprise $<$0.1\% of the total Encode
region, and were left out of the remaining analyses.

\begin{figure}

\includegraphics{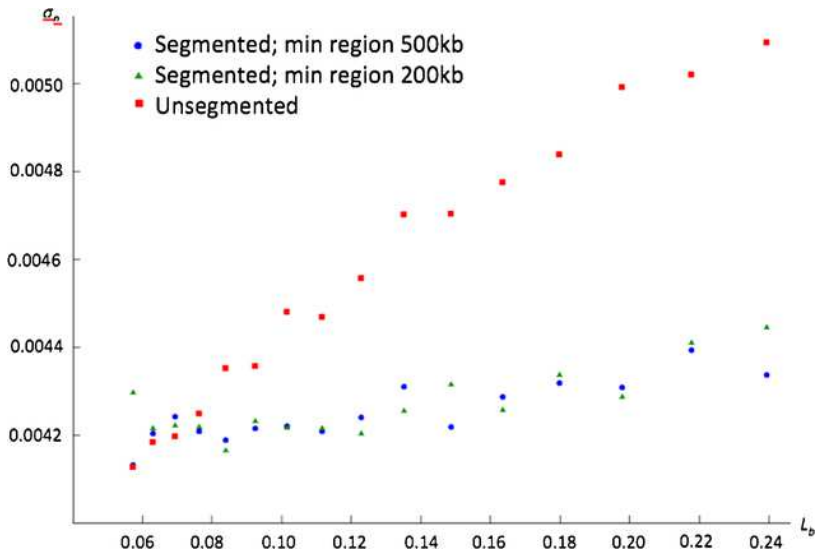}

\caption{Estimated $\sigma_n$ as a function of $L_b$ for 10,000
samples.}\label{fig:real:1}
\end{figure}

If the sequence were sufficiently homogeneous, we could forgo the
initial segmentation step. Figure \ref{fig:real:1} shows an estimate
of variance of $B_n$ (with the appropriate renormalization) for a reasonable
range of $L_b$, both before
and after segmentation.
Two trends are clearly evident. First, segmentation greatly reduces
the estimated variance. As we discussed in Section
\ref{secinconsist}, inhomogeneity of the sequence causes an inflated
estimate of variance. If the data were homogeneous, segmentation
should not change the variance estimate. Thus, the fact that the
estimated variances drop after segmentation for such a large range
of $L_b$'s suggests that the data are inhomogeneous. Second, and
more importantly, the estimated variance of $B_n$
increases sharply with increasing $L_b$ in the unsegmented data. This is
evidence of inhomogeneity in the mean of $B_n$ across this ENCODE region:
underlying shifts in mean, if ignored, can be mistaken for spurious
long range autocorrelation, which also implicitly runs against our
assumption. In either case, as Theorem \ref{SegBlock} suggests, we would be
overly conservative. Thus, a preliminary exploration of the data
convinces us that this ENCODE region is inhomogeneous in $I$ and/or $J$ and
segmentation is necessary.

%f5 ###

\begin{figure}

\includegraphics{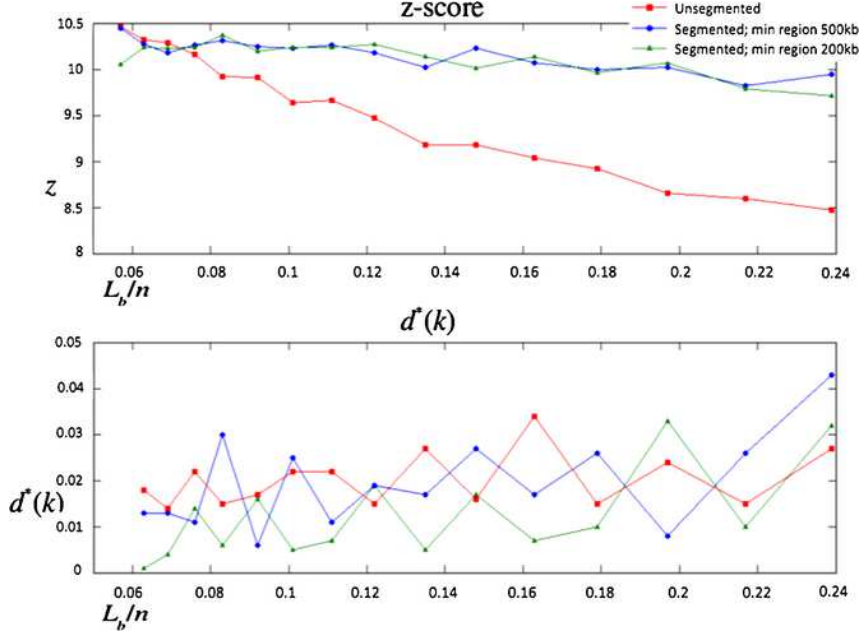}

\caption{Comparison of block subsampling distributions,
$\rho^{\beta} n$ vs. $\rho^{\beta+1}n$ under the IQR statistic.
Estimates $\hat{\sigma}_n$ and resulting $z$-\textit{scores} of $B_n$
shown.}\label{table:3}
\end{figure}

We found that 200 and 500~Kb gave 5 and 3 segments respectively.
Figure \ref{table:3} gives the results for 500~Kb. What is fairly
surprising, but
reassuring, is that over the whole broad range of $L_b$ considered,
the estimated SD of the statistic under the null was essentially
flat after segmentation. Flat here means that variability was
within a Monte Carlo SD for the 10,000 replications we used. We
would expect longer values of $L_b$ to include, in our estimate of
$\sigma$, additional covariance between distant genomic positions
captured by the extended block-length. The fact that this, by and
large, does not appear to be happening is consistent with our
hypothesis that the relevant mixing distance is indeed quite small
compared to the size of approximately stationary regimes.

We found that there is still moderate deviation from Gaussianity
in both the segmented and unsegmented case for $0.05<L_b<0.25$, both
in the tails, as detected by the Shapiro--Wilk test, and in the body
of the distribution under the Lilliefors test. With a sample size of 100,
neither test detects this departure, but at a sample size of only 500, it
is detected under a number of parameterizations of $L_b$. As we
discussed in
Section \ref{secchoosel}, the definition of stationarity depends on
the scale at which we view the genome. This suggests that our
segmentation still does not take care of inhomogenity in the
variance. Hence, as we have mentioned, if we use the variance for
the Gaussian approximation, our results are still conservative.

The scientific conclusion of this example is that, indeed, there is
strong association since the $z$-value is over 9 SDs. We note that the
effect of segmentation on our scientific conclusion is essentially
nonexistent. However, it is comforting to note that the change in
(with and without segmentation)
variance is in the correct direction.

%f6 ###

\subsection{The association of copy number variation with RefSeq
annotated exons in the human genome}
\label{real2}

%The ENCODE example came from a small genomic region, and represents
%a practical limit of our method: if the features of interest are not
%defined on sufficiently long genomic segments, block-sampling of the
%type we propose is impossible. Here we turn to an application of
%our method to a more suitable regime.

In this example, we reanalyze a published data set; this
reanalysis leads to a different conclusion from the one made by the
original paper. In 2006, Redon et al. published a set of 1445
genomic regions with observed Copy Number Variation (CNVs) across
individuals. These regions consist of both deletions and insertions,
and more than half of them overlap genes. In the paper, the authors
reported, among other things, a paucity of overlap with RefSeq
genes at a significance level of 0.05. The statistic that they used
is precisely our marginal formulation of the region overlap statistic
$R_n$, but the null
distribution to which they referred it is quite different. Their null
was computed by randomly permuting both genes and CNVs, and hence
treats the entire genome (or at least entire chromosomes) as
homogeneous, and the distances between feature-instances as
exponential. Thus, if feature-instance lengths were all 1~bp, this
would be a Poisson process. As discussed in Section~\ref{secsimIIb},
under our model this procedure provides an unbiased estimate of the
mean in the case of the $B_n$, but is unpredictable with respect to
its estimate of the variance. In the case of $R_n$, it is
unpredictable with respect to both the mean and the variance. Here,
for comparison with the result of Redon et al. (\citeyear{2006redon}), we examine
only $R_n$.

Although we have attempted to replicate this portion of the Redon
study, undoubtedly there are small differences between our efforts
and those of Redon et al. (\citeyear{2006redon}). For instance, we have masked all
genomic repeats in the ``Repeat Masker'' track on the UCSC genome
browser (\href{http://www.genome.ucsc.edu}{genome.ucsc.edu}). Redon et al. also considered patterns of
repeats in their analysis, but may have utilized an at least
slightly different map of genomic repeats. We find that $61.8\%$ of
the CNVs overlap RefSeq genes by at least 1 basepair. That is, we
wish to assess the significance of our observed statistic $R_n =
0.618$.

The calibration of the subsampling procedure is nontrivial,
especially in this application where we must consider the additional
parameter $L_r$. Hence, in the following we provide complete detail
regarding the calibration of our method for the data of Redon et al.
(\citeyear{2006redon}).

As before, our analysis begins with an assessment of the need for
segmentation. In this case, we are dealing with whole human
chromosomes, we expect that, in general, at least some segmentation is
necessary.
We segmented down to a minimum
segment length of 10,000,000~bps (10~Mbs), letting $L_s = 10$~Mb. The
mean length of these CNVs is around 250~Kb, and they are not
uniformly distributed, so we are compelled not to segment down to
regions much smaller than 10~Mb by our desire to capture the appropriate spatial
distribution of clusters of feature-instances. To assess
the sufficiency of the resulting segmentation, we examine the Gaussianity
of the segmented subsampling distributions. This examination is
tied to our selection of block length.

%f7 ###
\begin{figure}

\includegraphics{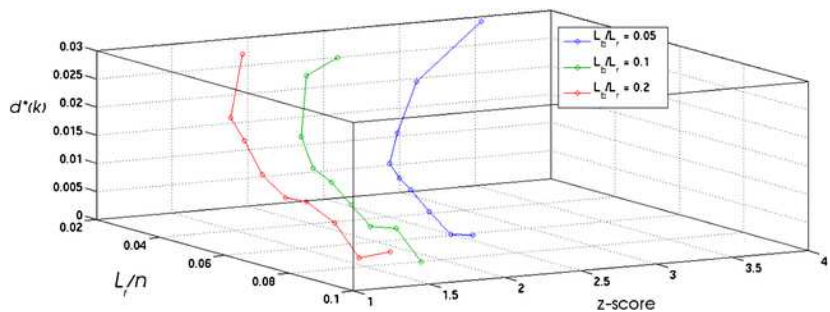}

\caption{The relationship between the estimated $z$-score, $d^*(k)$,
$L_r$ and $L_b$. As $L_r$ increases, our estimate of $\hat{\sigma}_n$
(not shown) increases, which drives the estimated $z$-score down. As
$L_r$ becomes too small, we lose the stability of our estimates, and
$d^*(k)$ increases. For the smallest value of $L_r$ shown here, the
estimated z-score increases sharply, but the corresponding value of
$d^*(k)$ indicates that this parameterization is unreliable. The ideal
parameterization under $d^*(k)$ is given by $L_r/n = 0.09$ and $L_b/L_r
= 0.20$.}
\label{fig:real:2}
\end{figure}

To select an inner block length, $L_b$, and an outer block-length,
$L_r$, we drew 10,000 samples for each of several lengths. We chose
to use a linear, rather than exponential, scale for $L_r/n$: we
selected 10 values from 0.01 to 0.10 in increments of 0.01. We
chose three values of $L_b/L_r$, 0.05, 0.10 and 0.20. Each of
these parameterizations yields several responses, including: an
estimated $z$-score, $d^*(k)$, and measures of Gaussianity. In Figure
\ref{fig:real:2}, we plot the relationship between the estimated
$z$-score, $d^*(k)$, $L_r$ and $L_b$. Regarding the Gaussianity of
the resulting distributions, at a significance level of 0.01 and a
sample size of 5000, neither the Shapiro--Wilk nor the Lilliefors
test rejected the null hypothesis of Gaussianity for any of the 30
explored parameterizations. To supplement our biological intuition
that segmentation is necessary when whole chromosomes are
considered, we used the same 30 parameterizations with the
unsegmented data, and performed the same tests to check the
Gaussianity of the resulting distributions. Of the 30
parameterizations, 3 showed departures from Gaussianity under
Lilliefors test, and 9 showed strong departures in the tails under
the Shapiro--Wilks test. This indicates, as expected, that
segmentation has substantially improved the Gaussianity of the
sample distributions. In practice, one might attempt a finer
segmentation in hopes of further reducing the (conservative) bias in
$\hat{\sigma}_n$. For this example we are satisfied with the current
segmentation.

The global minimum of $d^*(k)$ occurs for $L_r/n = 0.09$ and
$L_b/L_r = 0.20$. This parameterization yields an estimated $z$-score
of 1.25 and, therefore, we conclude that we cannot corroborate the
result of Redon et al. (\citeyear{2006redon}). Under our model it appears that CNVs
are, if anything, very slightly positively associated with genes ($p
\approx0.105$). We note that a few parameterizations, as shown in
Figure \ref{fig:real:2}, do produce $z$-scores greater than 2.
However, these parameterizations correspond to large values of
$d^*(k)$ and, furthermore, significance is in the opposite direction
reported by Redon et al. (\citeyear{2006redon}). This highlights the need for
carefully defined null distributions in genomic studies. We are not
suggesting that the results presented necessarily invalidate the
corresponding result of Redon et al. (\citeyear{2006redon}), but rather we caution
that scientific conclusions of this kind are predicated on how the
researcher defines ``at random,'' and that this definition should be
made to reflect, as much as possible, that which is known about the
actual distribution of genomic elements. We presume that authors
wish, in general, to err on the side of caution, and hence do not
wish to report significant association when the association can be
explained simply by a conservative choice of null.

\begin{supplement} [id-suppA]
\stitle{Some theorems in subsampling methods for genomic inference}
\sdescription{In Supplementary Material, we provide theoretical proofs to the theorems presented in the main text.}
\slink[doi]{10.1214/10-AOAS363SUPP}
\slink[url]{http://lib.stat.cmu.edu/aoas/363/supplement.pdf}
\sdatatype{.pdf}
\end{supplement}

\printaddresses

\end{document}